\documentclass[11pt,letterpaper]{article}
\pdfoutput=1
\usepackage{jcappub}

\usepackage{tabularx}
\usepackage{amsmath}
\usepackage{amsfonts}
\usepackage{amssymb}
\usepackage{mathrsfs}
\usepackage{graphicx}
\usepackage{color}
\usepackage[utf8]{inputenc}
\usepackage{amsfonts}
\usepackage{graphicx}
\usepackage{booktabs}
\usepackage{multirow}
\usepackage{makecell}
\usepackage{siunitx}
\usepackage{ragged2e}
\usepackage{rotating}
\usepackage{float}
\usepackage[dvipsnames]{xcolor}
\definecolor{darkred}{rgb}{0.5,0,0}
\definecolor{darkblue}{rgb}{0,0,0.5}
\definecolor{firebrick}{rgb}{0.75,0.125,0.125}
\definecolor{darkgreen}{rgb}{0,0.5,0}
\hypersetup{urlcolor=darkblue,
	    citecolor=darkgreen,
	    linkcolor=firebrick}

\usepackage{orcidlink}
\usepackage{lipsum}
\usepackage[normalem]{ulem}
\usepackage{enumitem}

\usepackage{subfigure}

\long\def\exclude#1{}

\newcommand{\ie}{{\it i.e.}}

\newcommand{\eg}{{\it e.g.}}

\newcommand{\eq}{Eq.}

\newcommand{\fig}{Fig.}

\newcommand{\Refe}{Ref.}
\newcommand{\Refes}{Refs.}
\newcommand{\equ}[1]{\eq~(\ref{equ:#1})}
\newcommand{\figu}[1]{\fig~\ref{fig:#1}}
\newcommand{\orcid}[1]{\href{https://orcid.org/#1}{\includegraphics[width=10pt]{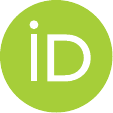}}}
\definecolor{forestgreen}{rgb}{0,0.6,0}

\graphicspath{figures/}

\title{Flavor Anisotropy in the High-Energy Astrophysical Neutrino Sky}

\author[a]{Bernanda Telalovic
\orcid{0000-0002-1406-502X}}
\emailAdd{bernanda.telalovic@nbi.ku.dk}

\author[a]{Mauricio Bustamante
\orcid{0000-0001-6923-0865}}
\emailAdd{mbustamante@nbi.ku.dk}

\affiliation[a]{Niels Bohr International Academy, Niels Bohr Institute,\\University of Copenhagen, DK-2100 Copenhagen, Denmark}

\date{\today}

\begin{document}

\abstract{
 High-energy astrophysical neutrinos, with TeV--PeV energies, offer unique insight into astrophysics and particle physics.  Their incoming directions and flavor composition---\ie, the proportion of $\nu_e$, $\nu_\mu$, and $\nu_\tau$ in their flux---are, individually, rewarding observables.  Combined, they offer new opportunities, hitherto unexplored, that we expose for the first time.  Anisotropy in the arrival directions of $\nu_e$, $\nu_\mu$, and $\nu_\tau$ may reveal multiple populations of neutrino sources, differently distributed in the sky, and test whether neutrinos of different flavor propagate preferentially along certain directions, such as expected from breaking Lorentz invariance.  Using 7.5 years of public IceCube High-Energy Starting Events, we make the first measurement of the directional flavor composition of high-energy astrophysical neutrinos, constrain the presence of flavor dipoles and quadrupoles, and improve constraints on ``compass asymmetries'' introduced by Lorentz-invariance violation.  In the near future, upcoming neutrino telescopes will improve these measurements across the board.
}

\maketitle


\section{Introduction}
Does the high-energy sky shine uniformly brightly in neutrinos of all flavors? In the last decade, the IceCube neutrino telescope discovered TeV--PeV astrophysical neutrinos~\cite{IceCube:2013cdw, IceCube:2013low, IceCube:2014stg, IceCube:2015gsk, IceCube:2015qii, IceCube:2016umi, IceCube:2020wum, IceCube:2021uhz}, bringing new insight to astrophysics~\cite{Anchordoqui:2013dnh, Ahlers:2018fkn, Ackermann:2019ows, Meszaros:2019xej, Halzen:2019qkf, Palladino:2020jol, AlvesBatista:2021eeu, Ackermann:2022rqc, Guepin:2022qpl} and fundamental physics~\cite{Gaisser:1994yf, Ahlers:2018mkf, Arguelles:2019rbn, Ackermann:2019cxh, AlvesBatista:2021eeu, Ackermann:2022rqc}.  Today, with a growing number of detected neutrinos and experimental advances, we are able to tackle the directional dependence of the fluxes of high-energy $\nu_e$, $\nu_\mu$, and $\nu_\tau$. This is an observable rich with potential, that will greatly benefit from the synergy between current and upcoming neutrino telescopes.

High-energy neutrino telescopes measure neutrino arrival directions.  Searches for astrophysical neutrino point sources use this to look for concurrent electromagnetic emission---which has revealed the first sources~\cite{IceCube:2018cha, IceCube:2018dnn, Stein:2020xhk, Reusch:2021ztx, IceCube:2022ham, IceCube:2022der}---clusters of neutrinos~\cite{IceCube:2015usw, Feyereisen:2016fzb, IceCube:2017der, IceCube:2019cia, Capel:2020txc, IceCube:2020nig, IceCube:2021xar}, and correlations with astrophysical catalogs~\cite{IceCube:2016ipa, Padovani:2016wwn, Moharana:2016mkl, IceCube:2016qvd, Resconi:2016ggj, Turley:2018biv, Lunardini:2019zcf, IceCube:2019cia, IceCube:2019xiu, Plavin:2020emb, Buson:2022fyf, Plavin:2022oyy, IceCube:2023htm, Buson:2023irp, Bellenghi:2023yza} and ultra-high-energy cosmic rays~\cite{Moharana:2015nxa, IceCube:2015afa, Resconi:2016ggj, IceCube:2022osb}.  Searches via the angular power spectrum of arrival directions---closer to our work---are rarer~\cite{IceCube:2014gax, IceCube:2014rqf, Leuermann:2016oxu, Fang:2016hyv, Dekker:2018cqu}, but may soon be bolstered.  Separately, neutrino telescopes measure the flavor composition of high-energy astrophysical neutrinos, \ie, the proportions of neutrinos of different flavor in the total flux~\cite{Mena:2014sja, Palomares-Ruiz:2015mka, IceCube:2015rro, Palladino:2015vna, IceCube:2015gsk, Vincent:2016nut, IceCube:2018pgc, IceCube:2020fpi}.  This is a versatile, if challenging, observable to study neutrino astrophysics~\cite{Rachen:1998fd, Athar:2000yw, Crocker:2001zs, Barenboim:2003jm, Beacom:2003nh, Beacom:2004jb, Kashti:2005qa, Mena:2006eq, Kachelriess:2006ksy, Lipari:2007su, Esmaili:2009dz, Choubey:2009jq, Hummer:2010ai, Winter:2013cla, Palladino:2015zua, Bustamante:2015waa, Biehl:2016psj, Bustamante:2019sdb, Bustamante:2020bxp, Song:2020nfh} and fundamental physics~\cite{Beacom:2002vi, Barenboim:2003jm, Beacom:2003nh, Beacom:2003eu, Beacom:2003zg, Serpico:2005bs, Mena:2006eq, Lipari:2007su, Pakvasa:2007dc, Esmaili:2009dz, Choubey:2009jq, Esmaili:2009fk, Bhattacharya:2009tx, Bhattacharya:2010xj, Bustamante:2010nq, Mehta:2011qb, Baerwald:2012kc, Fu:2012zr, Pakvasa:2012db, Chatterjee:2013tza, Xu:2014via, Aeikens:2014yga, Arguelles:2015dca, Bustamante:2015waa, Pagliaroli:2015rca, deSalas:2016svi, Gonzalez-Garcia:2016gpq, Bustamante:2016ciw, Rasmussen:2017ert, Dey:2017ede, Bustamante:2018mzu, Farzan:2018pnk, Ahlers:2018yom, Brdar:2018tce, Palladino:2019pid, Ahlers:2020miq, Karmakar:2020yzn, Fiorillo:2020gsb, Song:2020nfh}.  

 \begin{figure}[h!]
 \centering
 \includegraphics[trim={0 8.5cm 0 0}, clip, width=0.7\columnwidth]{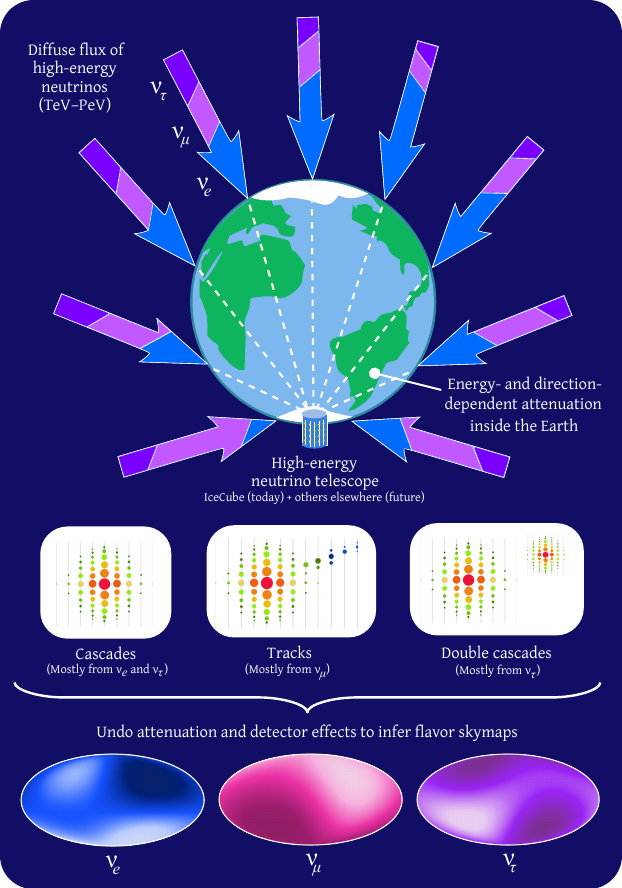}
 \caption{\textbf{\textit{Inferring the directional dependence of high-energy astrophysical neutrino diffuse flux of each flavor.}}  We infer the directional distributions of the underlying skymaps of $\nu_e$, $\nu_\mu$, and $\nu_\tau$ from the directional distribution of different types of events (cascades, tracks, double cascades). }
 \label{fig:schematic}
\end{figure}

The capabilities to infer neutrino arrival directions and flavor composition have remained largely separate, as flavor composition has been measured only averaged across the sky.  This has been motivated by the standard expectation that the neutrino sources, and their yields of $\nu_e$, $\nu_\mu$, and $\nu_\tau$, are distributed roughly isotropically, and that the flavor oscillations evolving those yields en route to Earth are unaffected by specific neutrino trajectories.  Yet, we show that there is latent potential, accessible already today, in combining those capabilities to probe the directional dependence of the flavor composition.  For astrophysics, flavor anisotropy could reveal multiple populations of sources, hosting different neutrino-production processes, distributed differently in the sky.  For particle physics, it could reveal preferred directions of propagation for neutrinos of different flavors, as expected in some forms of Lorentz-invariance violation. These are opportunities for progress in unexplored directions.
For the first time, we explore this potential, using present IceCube data~\cite{IceCube:2020wum} and {near-future} projections.

Figure~\ref{fig:schematic} sketches our procedure.  From a sample of events of different types detected by IceCube and future detectors---cascades, tracks, and double cascades---we infer the angular distributions of the diffuse fluxes of $\nu_e$, $\nu_\mu$, and $\nu_\tau$.  There are two main challenges.  First, an event of a certain type may have been produced by {any} neutrino flavor.  We account for this by using a detailed description of the detector response.  Second, the angular uncertainty of many of {the events} is large, up to tens of degrees.  This limits our search to large-scale angular structures---dipoles and quadrupoles. Yet, even at these scales, we can start probing mechanisms that can introduce neutrino flavor anisotropy. 

We demonstrate this utility by using the directional flavor composition recovered from IceCube's High Energy Starting Events (HESE)~\cite{IceCube:2013low, IceCube:2014stg} dataset to obtain preliminary constraints on Lorentz-invariance violating effects.

\section{Astrophysical neutrino production}

Astrophysical processes that involve charged particle acceleration produce neutrinos as byproducts. Both the flavor composition and the spectral shape of the neutrino flux depend are governed at least in part by the production mechanisms. Currently, astrophysical neutrino sources and the mechanisms that produce high energy neutrinos are only beginning to be discovered. 

\subsection{Neutrino sources}
The TeV--PeV astrophysical neutrinos seen by IceCube are presumably produced in interactions between cosmic-ray protons of tens of PeV and matter~\cite{Margolis:1977wt, Stecker:1978ah, Kelner:2006tc} or radiation~\cite{Stecker:1978ah, Mucke:1999yb, Kelner:2008ke, Hummer:2010vx, Morejon:2019pfu} inside cosmic accelerators, predominantly extragalactic~\cite{Hillas:1984ijl, Anchordoqui:2018qom, AlvesBatista:2019tlv, Ackermann:2022rqc}.  These interactions produce high-energy pions that, upon decaying, produce high-energy neutrinos, \ie, $\pi^+ \to \mu^+ + \nu_\mu$, followed by $\mu^+ \to e^+ + \nu_e + \bar{\nu}_\mu$, and their charge-conjugated processes.

From the full pion decay chain, our nominal expectation~\cite{Bustamante:2015waa} is that neutrinos leave their sources (S) with flavor composition $(f_{e, {\rm S}}, f_{\mu, {\rm S}}, f_{\tau, {\rm S}}) = \left(\frac{1}{3}, \frac{2}{3}, 0\right)$, where $f_{\alpha, {\rm S}}$ is the fraction of $\nu_\alpha + \bar{\nu}_\alpha$ emitted ($\alpha = e, \mu, \tau$).  In sources {with} intense magnetic fields, the intermediate muons might cool via synchrotron radiation, yielding the ``muon-damped'' case, $(0,1,0)_{\rm S}$.  Neutron decays yield a pure-$\bar{\nu}_e$ flux, $(1,0,0)_{\rm S}$.  {There are other possibilities}~\cite{Razzaque:2004yv, Ando:2005xi, Kashti:2005qa, Kachelriess:2006ksy, Kachelriess:2007tr, Hummer:2010ai, Bustamante:2015waa, Carpio:2020app, Arguelles:2023bxx}, but we adopt these three as benchmarks. Other production channels contribute, especially at high energies~\cite{Mucke:1999yb, Hummer:2010vx, Morejon:2019pfu}.

En route to Earth, neutrino oscillations~\cite{Pontecorvo:1967fh, Super-Kamiokande:1998kpq, SNO:2002tuh} transform the flavor composition into $f_{\alpha, \oplus} = \sum_{\beta=e,\mu,\tau} P_{\beta\alpha} f_{\beta, {\rm S}}$, where $P_{\beta\alpha} = \sum_{i=1}^3 \lvert U_{\beta i} \rvert^2 \lvert U_{\alpha i} \rvert^2$ is the average $\nu_\beta \to \nu_\alpha$ flavor-transition probability, and $U$ is the lepton mixing matrix~\cite{ParticleDataGroup:2022pth}.  We evaluate it using the present-day best-fit values of the neutrino mixing parameters~\cite{Esteban:2020cvm}, ignoring uncertainties whose effect on $f_{\alpha, \oplus}$ will be tiny by 2030~\cite{Song:2020nfh}. The nominal expectation, from pion decay, is about $\left( 0.30, 0.36, 0.34 \right)_\oplus$.

\subsection{Sources of background}
For neutrinos in the TeV--PeV energy range observable by IceCube-like detectors, the irreducible sources of background that contribute to the event rates are produced in cosmic-ray interactions in the atmosphere: muons ($\mu$), conventional (c) neutrinos from pion decays, and prompt (pr) neutrinos from kaon decays. In our analysis, we allow the normalizations of their fluxes to be free parameters, and we rely on the IceCube HESE Monte Carlo (MC) for the detailed modeling of their energy and directional distributions.


\section{Neutrino detection and flavor inference}

\subsection{Current detection with IceCube}

IceCube is the largest neutrino telescope in operation.  It instruments around 1~km$^3$ of Antarctic ice with photomultipliers that capture the Cherenkov light emitted by particle showers from neutrino interactions in the ice.  At TeV--PeV energies, a neutrino typically undergoes deep inelastic scattering~\cite{IceCube:2017roe, Bustamante:2017xuy, IceCube:2018pgc, IceCube:2020rnc} on a nucleon $N$, either charged-current (CC), $\nu_\alpha + N \to l_\alpha + X$, where $X$ represents hadrons, or neutral-current (NC), $\nu_\alpha + N \to \nu_\alpha + X$.  Outgoing charged particles shower and radiate.  

We focus on the IceCube HESE data~\cite{IceCube:2013low, IceCube:2014stg}, in which the neutrino interaction occurs inside the detector volume.  HESE samples are dominated by astrophysical neutrinos; contamination from atmospheric neutrinos is small due to a self-veto~\cite{Schonert:2008is, Gaisser:2014bja}. HESE events have a resolution on their energy, $E_{\rm evt}$, of about 10\% in $\log_{10}(E_{\rm evt}/{\rm GeV})$~\cite{IceCube:2013dkx}, and varying angular resolution.  From the amount of light collected, and its spatial and temporal distributions, analyses infer the energy and direction of the parent neutrino.

There are three types of events that IceCube can detect at these energies---tracks, cascades, and double cascades (\figu{schematic})---from which we infer the flavor composition.

Tracks are made by $\nu_\mu$ CC interactions and by 17\% of $\nu_\tau$ CC interactions where the outgoing tau decays into a muon~\cite{ParticleDataGroup:2022pth}.  The muon leaves an identifiable km-long track of light in its wake.  Tracks have typical angular resolution of about $1^\circ$~\cite{IceCube:2021oqo}.

Cascades are made by all other interactions.  In $\nu_e$ CC interactions, the showers initiated by the outgoing electron and hadrons combine.  In $\nu_\tau$ CC interactions, the outgoing tau decays promptly and its shower combines with that of the hadrons (when the tau does not decay into a muon).  In NC interactions of all flavors, only the outgoing hadrons shower.  Cascades have typical angular resolution of tens of degrees~\cite{IceCube:2013dkx}.

Double cascades are made by $\nu_\tau$ CC interactions when two cascades are seen in tandem: from $\nu_\tau N$ scattering and, later, from the decay of the tau into hadrons~\cite{Learned:1994wg}.  They have typical angular resolution of 30$^\circ$~\cite{IceCube:2020fpi}.

Hence, it is generally not possible to infer neutrino flavor event-by-event.  The mis-identification of tracks, cascades, and double cascades complicates it further~\cite{Mena:2014sja, IceCube:2014rwe, Palomares-Ruiz:2015mka, IceCube:2020wum, IceCube:2020fpi}.  Instead, analyses, including ours, infer the flavor composition of a sample of events~\cite{Mena:2014sja, Palomares-Ruiz:2015mka, IceCube:2015rro, Palladino:2015vna, IceCube:2015gsk, Vincent:2016nut, IceCube:2018pgc, IceCube:2020fpi}. IceCube data alone faces the issue of the Earth's volume partially blocking out the flux of astrophysical neutrinos coming into the telescope from the northern hemisphere. We will later outline how we use the Monte Carlo associated to the HESE data release to compensate for this effect. 

\subsection{Upcoming neutrino telescopes}\label{sec:future_detectors}

In the future, IceCube will not be the only neutrino telescope that will be able to infer high-energy neutrino flavor and arrival directions. Combining data from all the current and upcoming collaborations will bolster both flavor reconstruction analyses, driven by the increase in statistics, as well as analyses focusing on neutrino arrival directions. In the latter case, combining exposures of several telescopes that are distributed across the globe has the added effect of compensating for the flux reduction in each individual telescope due to the Earth. Furthermore, some of these upcoming telescopes, \eg, KM3NeT, are water-based Cherenkov detectors, whose angular resolution is predicted to be better than in ice at these neutrino energies, which would further bolster directional searches. 

\begin{figure}[h!]
 \centering
 \includegraphics[width=0.8\textwidth]{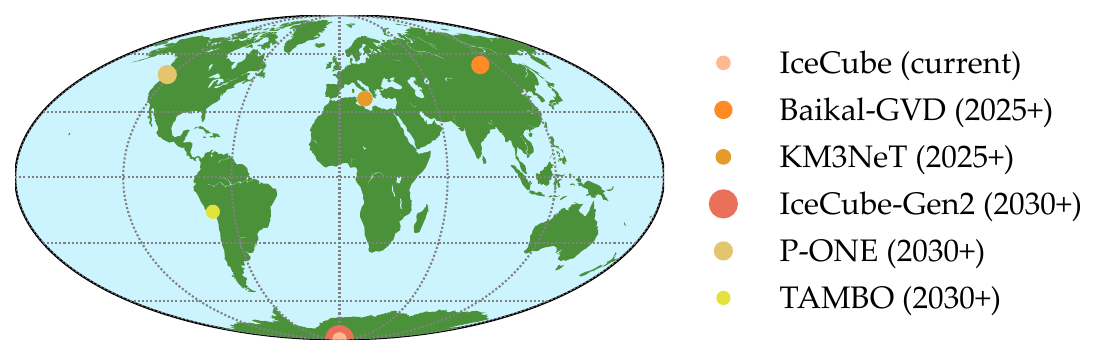}
 \caption{\textbf{\textit{Locations of neutrino telescopes used for forecasting.}} The marker size represents the size of the detector relative to the size of IceCube.}
 \label{fig:future-detectors}
\end{figure}

Figure~\ref{fig:future-detectors} and Table~\ref{tab:detectors} show the positions, sizes, and dates of start and end of operations of the neutrino telescopes that we use to produce our results: IceCube, today, and Baikal-GVD, IceCube-Gen2, KM3NeT, P-ONE, and TAMBO, in coming years.  Save for TAMBO, these are all in-water and in-ice Cherenkov detectors.  (We do not consider HUNT~\cite{Huang:2023mzt} or TRIDENT~\cite{Ye:2022vbk}, but future analyses could.)  For upcoming detectors, we only consider the start dates of their planned final configurations, as in \Refes~\cite{Song:2020nfh, Fiorillo:2022rft, Arguelles:2023bxx}, even if they may be running in partial configurations earlier; this entails only a small loss in cumulative number of detected events.

\begin{table}[t!]
\caption{\label{tab:detectors}\textbf{\textit{High-energy neutrino telescopes used for forecasting.}} The detector effective areas are given in fractions of the effective volume of IceCube (IC).  This factor rescales the IceCube HESE Monte Carlo sample.  The start and end years are estimated.  For IceCube, the start year is for the final detector configuration and the end year is when IceCube-Gen2 is estimated to start operations.}
    \centering
    \begin{tabular}{ccccc}
         \multirow{2}{*}{Detector} & \multirow{2}{*}{\makecell{Position (lat, lon)}} & \multirow{2}{*}{\makecell{Effective area}} & \multirow{2}{*}{\makecell{Start year}} & \multirow{2}{*}{\makecell{End year}} \\[10pt]
   \hline\\[-10pt]
   IceCube & ($0^\circ$, $-90^\circ$) & 1 IC & 2013 & 2030 \\[2pt]
   Baikal-GVD & ($108.17^\circ$, $53.56^\circ$) & 1.5 IC & 2025 & $\cdots$ \\[2pt]
   KM3NeT & ($16.1^\circ$, $36.27^\circ$) & 2.8 IC & 2025 & $\cdots$ \\[2pt]
   IceCube-Gen2 & ($0^\circ$, $-90^\circ$) & 8 IC & 2030 & $\cdots$ \\[2pt]
   P-ONE & ($-123.36^\circ$, $48.43^\circ$) & 3.2 IC & 2030 & $\cdots$ \\[2pt]
   TAMBO & ($-15.6^\circ$, $-71.89^\circ$) & 0.5 IC & 2030 & $\cdots$
    \end{tabular}
    \label{tab:my_label}
\end{table}

\smallskip

\textbf{\textit{Baikal-GVD}}~\cite{Dvornicky:2023hih}, the successor of Baikal NT-200~\cite{BAIKAL:1997iok}, is an in-water Cherenkov detector in Lake Baikal, Russia, currently under construction.  Since 2018, it has been operating in partial configuration; in 2022, its size was about $0.35$ km$^3$.  Already, it has reported the detection of a high-energy neutrino~\cite{Baikal-GVD:2022fmn} from the blazar TXS 0506+056 previously observed by IceCube~\cite{IceCube:2018cha, IceCube:2018dnn}, and, tentatively, of the diffuse flux of high-energy astrophysical neutrinos seen by IceCube~\cite{Baikal-GVD:2022fis}.  We assume a start date of the full Baikal-GVD of 2025, with an effective volume of $1.5$ km$^3$.

\smallskip

\textbf{\textit{KM3NeT}}~\cite{KM3Net:2016zxf}, the successor of ANTARES~\cite{ANTARES:2011hfw}, is an in-water Cherenkov detector in the Mediterranean Sea, currently under construction.  ARCA, the detector sub-array dedicated to high-energy neutrinos, had 19 of the planned 115 units deployed in 2022.  Because the scattering length of Cherenkov light in liquid water is longer than in ice, the direction of detected events---including cascades---can be reconstructed more accurately in KM3NeT than in IceCube~\cite{KM3Net:2016zxf}.  We account for this by improving the angular resolution of each cascade in the HESE MC by a factor of 3, which conservatively approximates expectations for KM3NeT.  We assume a start date of the full KM3NeT of 2025, with an effective volume of 2.8~km$^3$.

\smallskip

\textbf{\textit{IceCube-Gen2}}~\cite{IceCube-Gen2:2020qha} is the envisioned successor of IceCube, located also at the South Pole.  Like IceCube, it is an in-ice Cherenkov detector.  It comprises 120 new detector strings that, among other improvements, will significantly enhance the measurement of the diffuse high-energy neutrino flux~\cite{Ackermann:2022rqc}.  (In addition, IceCube-Gen2 includes an array of radio antennas for the discovery of ultra-high-energy neutrinos, of EeV-scale energies, which we do not account for here; see, \eg, \Refes~\cite{Valera:2022ylt, Fiorillo:2022ijt, Valera:2022wmu, Valera:2023ayh}.)  We assume a start date for the full IceCube-Gen2 of 2030, with an effective volume of 8~km$^3$.

\smallskip

\textbf{\textit{P-ONE}}~\cite{P-ONE:2020ljt, Haack:2023uwd} (the Pacific Ocean Neutrino Experiment) is an envisioned in-water Cherenkov detector, to be located in the Cascadia Basin, Canada.  It is presently in the prototype stage, with ongoing development of the first prototype string, and planned deployment of the first cluster of strings in 2024--2025~\cite{Henningsen:2023slc}.  P-ONE is also a driving force of PLE$\nu$M, the Planetary Neutrino Monitoring System~\cite{Schumacher:2021hhm, Haack:2023uwd}.   We assume a start date of the full P-ONE of 2030, with an effective volume of 3.2~km$^3$.

\smallskip

\textbf{\textit{TAMBO}}~\cite{Romero-Wolf:2020pzh, TAMBO:2023plw} (the Tau Air-Shower Mountain-Based Observatory) is an envisioned detector of high-energy $\nu_\tau$, to be located in the Peruvian Andes.  Unlike the other detectors that we consider, TAMBO consists of a surface array of Cherenkov-detector tanks deployed on a canyon wall.  They detect the particle showers triggered by tauons decaying in air, after being produced by Earth-skimming $\nu_\tau$ that interact on the opposite canyon wall~\cite{Fargion:1999se}.  Since TAMBO is sensitive only to $\nu_\tau$, we model its response by only keeping the HESE MC events that originate from a $\nu_\tau$ interaction and yield a double cascade.  This is admittedly a rough approximation, and should be improved in the near future~\cite{Lazar:2023fgq}. However, given the small size of TAMBO relative to the other detectors, the mismodeling has limited impact on our results.  We assume a start date of the full TAMBO of 2030, with an effective volume of 0.5~km$^3$.


\section{Directional flavor flux modeling}
\subsection{The astrophysical neutrino flavor flux}

Interactions energetic enough to produce neutrinos in the TeV--PeV range, via processes described above, can occur in a multitude of astrophysical sources that, together, contribute to the diffuse astrophysical neutrino flux we observe at Earth. Because of the paucity of present-day HESE data, we group detected events in a single energy bin, \ie, from each direction, we only use the total number of events of each type.  Accordingly, we take the flavor composition of the astrophysical neutrinos to be independent of neutrino energy.  To ease comparison between present-day and projected results, we maintain this choice even in our projections, but point out that future, larger event rates could render the use of multiple energy bins feasible. However, as different types of event rates are dependent on the spectral shape of the neutrino flux, we still have to account for the energy dependence in our flux model. Over the whole HESE energy range, the cpresent-day data are consistent with the energy spectrum of high-energy astrophysical neutrinos being a power law in neutrino energy. We keep this model throughout our analysis, parameterized by a spectral index, which is equal for each flavor, \ie, the fluxes of each neutrino flavor at Earth are all described by this power-law spectral index.

To introduce directional dependence into the fluxes of each flavor, we tessellate the sky into $N_{\rm pix} = 12$ equal-area pixels using HEALPix, with parameter $N_{\rm side} = 1$~\cite{Gorski:2004by, healpix_url}. We divide the all-flavor flux normalization equally in all 12 pixels, and parameterize the flavor composition in each pixel by introducing a pixel-wise independent flavor ratio $f_{\alpha,i}$ ($\alpha = e,\mu,\tau$ and $i = 1,\cdots,12$), where $f_{e,i}+f_{\mu,i}+f_{\tau,i} = 1$ in each pixel. In this way we build three flavor flux skymaps, and generate three IceCube event type, $t$, skymaps: cascades (c), tracks (tr), and double cascades (dc). Because IceCube cannot distinguish between events triggered by $\nu_\alpha$ and $\bar{\nu}_\alpha$, we assume their fluxes to be equal.  However, we reweigh the MC samples separately for $\nu_\alpha$ and $\bar{\nu}_\alpha$; their MC weights are similar, except in the case of $\nu_e$ {vs.}~$\bar{\nu}_e$, where only the latter has contributions from the Glashow resonance~\cite{Glashow:1960zz, IceCube:2021rpz} at multi-PeV energies. In this way, even though we assume their underlying fluxes are equal, the potentially different $\nu$ vs. $\bar{\nu}$ contributions to the event rates in each pixel, and of each type are still accounted for. Overall, we adopt the following flavor flux model:
\begin{equation}
 \label{equ:flav_flux_earth}
 \Phi_{\nu_\alpha}
 (E, \theta_z, \phi, \boldsymbol{\theta})
 =
 \left( \frac{\Phi_0}{N_{\rm pix}} \right)
 \left(\frac{E}{100~{\rm TeV}}\right)^{-\gamma} 
 \frac{f_{\alpha, \oplus} (\theta_z, \phi)}{2} \;,
\end{equation}
where $\Phi_0$ is the all-flavor flux normalization, $N_\text{pix}=12$ is the number of directional bins on the sky, $\gamma$ is the power-law spectral index, and the directional flavor ratio of $\nu_\alpha$, $f_{\alpha, \oplus} (\theta_z, \phi)$, is determined by the polar and azimuthal angles, $(\theta_z, \phi)$. We say that $f_{\alpha, \oplus} (\theta_z, \phi) = f_{\alpha, i}$ when the pixel $i$ contains the direction $(\theta_z, \phi)$, \ie, we assume that the flavor composition inside any particular pixel is constant inside that pixel, though different pixels can have different flavor composition. The factor of 2 in the denominator divides the flux of $\nu_\alpha + \bar{\nu}_\alpha$ evenly between $\nu_\alpha$ and $\bar{\nu}_\alpha$.  In this case, $\boldsymbol{\theta} \equiv (\Phi_0, \gamma, \left\{f_{\alpha, i} \right\}_{i=1}^{N_{\rm pix}=12})$ are the flux model parameters, their values determined in fits to data.

\subsection{Calculating event rates}

To perform a directional flavor inference on both the real and simulated data, we need to use our flux model to determine the event rates seen in the detector(s). To do this, we  account for uncertainties arising from directional and energy reconstruction, each detector's field of view, and inherent uncertainties in the flux model parameters. In addition, it is difficult to infer each event's flavor directly, as the event types are only statistically correlated with the interacting neutrino's flavor, also in an energy-dependent way. Currently, the most accurate modeling of the detector's response to flavor identification, field of view, and energy reconstruction, is captured in the HESE MC. We therefore use it to predict the event rates for our statistical analysis. With this approach, we are implicitly including the average uncertainty in directional reconstruction for each event, modeled as a 2D normal distribution. 

However, the analysis of the 7.5-year HESE sample by the Collaboration~\cite{IceCube:2020wum} reported more detailed probability distributions of the reconstructed directions of the 102 detected events, each parametrized by a Fisher-Bingham distribution that is flexible enough to capture the variety and asymmetry of the directional uncertainty. Such distributions are not provided for the events in the HESE MC sample, on account of them being computationally taxing to produce; for them, only the mean reconstructed directions are provided. Because of this, we opt for an Asimov analysis that uses the mean reconstructed directions of present and projected HESE samples.  Including the uncertainty in reconstructed directions would likely weaken the significance of our claims, though it is not possible to say by how much without dedicated analysis. 

Our analysis has a few differences compared to the analysis of the 7.5-year HESE sample performed by the IceCube Collaboration~\cite{IceCube:2020wum}.  The primary difference is that the IceCube analysis assumed the flavor composition at Earth to be isotropic and equal to $\left(\frac{1}{3}, \frac{1}{3}, \frac{1}{3} \right)_\oplus$, unlike ours.  Beyond that, our chosen priors on $\Phi_0$, $\gamma$, and $\Phi_{\rm pr}$ are more restrictive; see Table IV.1 in \Refe~\cite{IceCube:2020wum}.  For these parameters, our priors are, in fact, motivated by the results found in that analysis.  In contrast, our priors on $\Phi_{\rm c}$ and $\Phi_\mu$ are the same as in \Refe~\cite{IceCube:2020wum}.  

Further, the IceCube analysis in \Refe~\cite{IceCube:2020wum} adopted a more sophisticated likelihood prescription~\cite{Arguelles:2019izp} and included more sources of systematic uncertainty than our analysis; see Table IV.1 in \Refe~\cite{IceCube:2020wum}.  Including these additional sources in our analysis would have significantly increased the size of the model parameter space to explore---which, for us, is already larger than for \Refe~\cite{IceCube:2020wum}---and would have demanded computational resources beyond the considerable ones already used in our analysis.  Our aim is to present directional flavor composition as a new  observable.  A measurement of it that includes all the sources of systematic uncertainty considered in \Refe~\cite{IceCube:2020wum} is best left to the Collaboration, which has a better handle on  detector details than what is publicly available~\cite{IC75yrHESEPublicDataRelease}.

In the following, we describe the procedure we employ for using the IceCube HESE Monte Carlo to account for the detector effects of IceCube itself, as well how we use it to estimate the responses of upcoming neutrino telescopes. 

\subsubsection{HESE MC and reweighing}\label{sec:mc_modifications}

The public MC sample of simulated IceCube HESE events~\cite{IC75yrHESEPublicDataRelease} was released together with the 7.5-year HESE sample~\cite{IceCube:2020wum}. For each simulated neutrino event, the MC contains its true flavor, direction and energy, as well as the event morphology its reconstructed as, the reconstructed energy and direction. In this way, it encodes the flavor, energy, and direction-dependent detector response for which we need to account in order to recover the flavor composition of the underlying flux. 
In addition, it also models the background contributions to the astrophysical component of the flux. For the true and reconstructed directional information, the MC contains only the zenith coordinate of each event, and not the azimuthal component. To compare it to data, which contains both the polar and azimuthal reconstructed directions for each event, we randomly assign an azimuthal ``true'' direction, and we compute the reconstructed azimuth angle of the event by randomly sampling a value from a normal distribution centered at the neutrino azimuth angle, and with a spread equal to the difference between the neutrino zenith angle and the reconstructed zenith angle of the event. As the HESE MC was generated under an isotropic flux assumption, this modification does not  introduce any directional biases when we use the MC in our statistical analyses.

\subsubsection{Astrophysical flux modeling with the MC}

The IceCube HESE MC sample was generated assuming the reference isotropic diffuse power-law flux
\begin{equation}
 \label{equ:phi_ref}
 \Phi_{\nu_x, {\rm ref}}(E)
 =
 \frac{\Phi_{0, {\rm ref}}}{6}
 \left(\frac{E}{100~{\rm TeV}}\right)^{-\gamma_{\rm ref}} \;,
\end{equation}
where the flux parameters are fixed to their best-fit values inferred from the 7.5-year IceCube HESE sample~\cite{IceCube:2020wum}, \ie, $\Phi_{0, {\rm ref}} = 5.68 
\cdot 10^{-18}$~GeV$^{-1}$~cm$^{-2}$~s$^{-1}$~sr$^{-1}$ and $\gamma_{\rm ref} = 2.89$.  The factor of 6 in \equ{phi_ref} divides the all-flavor flux evenly among $\nu_e$, $\nu_\mu$, $\nu_\tau$, $\bar{\nu}_e$, $\bar{\nu}_\mu$, and $\bar{\nu}_\tau$.  This implicitly assumes that the flux is equally divided between neutrinos and antineutrinos; we keep the same assumption when testing other astrophysical neutrino fluxes below, but allow the flavor composition to change.

From the HESE MC sample~\cite{IC75yrHESEPublicDataRelease}, we extract the weight, $w_k^{t, \alpha}$, associated to the $k$-th event of topology $t$ generated by a $\nu_\alpha$ of energy $E_k$, zenith angle $\theta_{z, k}$, and azimuth angle $\phi_k$. Similarly, the weight for the event generated by a $\bar{\nu}_\alpha$ we label as $\bar{w}_{k}^{t, \alpha}$.  We use the weights to transform the event distribution of MC events for different choices of diffuse neutrino fluxes, $\Phi_{\nu, \alpha} = \Phi_{\bar{\nu}, \alpha}$, including those with a directional dependence.  The mean expected number events of type $t$, due to astrophysical $\nu_\alpha$ plus $\bar{\nu}_\alpha$, detected inside the $i$-th skymap pixel, after exposure time $T$, is
\begin{equation}
 \label{equ:mc_reweigh_astro}
 N_{{\rm ast}, i}^{t,\alpha}(\boldsymbol{\theta})
 =
 \sum_{\text{event }k\text{ in pixel }i}
 \frac{\Phi_{\nu_\alpha}(E_k, \theta_{z, k}, \phi_k, \boldsymbol{\theta}) T}
 {\Phi_{\nu_\alpha, {\rm ref}}(E_k) T_{\rm ref}}
 \left( w_k^{t, \alpha} + \bar{w}_k^{t, \alpha} \right) \;,
\end{equation}
where the sum includes only events whose reconstructed positions are within that pixel, $\boldsymbol{\theta}$ represents all of the parameters on which the test neutrino flux depends (more on this later), and $T_{\rm ref} = 2635$~days is the exposure time of the IceCube HESE analysis~\cite{IceCube:2020wum}. 

\subsubsection{Background contributions with the MC}

In addition to events from high-energy astrophysical neutrinos, we account for events due to the irreducible background of atmospheric neutrinos and muons. From the IceCube HESE MC, adjusted to include also azimuth angles, we extract the baseline number of events detected from conventional atmospheric neutrinos, $N_{{\rm c}, i}^t$, prompt atmospheric neutrinos, $N_{{\rm pr}, i}^t$, and atmospheric muons, $N_{\mu, i}^t$.  

For the conventional atmospheric neutrino flux, the baseline prescription, from \Refe~\cite{Honda:2006qj}, is obtained using the modified DPMJET-III generator~\cite{Roesler:2000he}.  For the prompt atmospheric neutrino flux, the baseline prescription is from \Refe~\cite{Bhattacharya:2015jpa}.  To date, there are only upper limits on this flux.  For the atmospheric muon flux, the baseline prescription is from air-shower simulations from {\tt CORSIKA}~\cite{Heck:1998vt}, using the Hillas--Gaisser H4a cosmic-ray flux model~\cite{Gaisser:2013bla} and the Sibyll~2.1 hadronic interaction model~\cite{Ahn:2009wx}. 

We maintain the baseline shapes of the energy distributions of the background events, but rescale their rates by allowing their flux normalization constants, $\Phi_{\rm c}$, $\Phi_{\rm pr}$, and $\Phi_\mu$, to float independently of each other in fits to data.  Thus, the total number of events due to atmospheric backgrounds is
\begin{equation}\label{eq:background_event_rates}
 N_{{\rm atm}, i}^t (\boldsymbol{\eta})
 =
 \Phi_{\rm c} N_{{\rm c}, i}^t
 +
 \Phi_{\rm pr} N_{{\rm pr}, i}^t
 +
 \Phi_\mu N_{\mu, i}^t \;,
\end{equation}
where $\boldsymbol{\eta} \equiv (\Phi_{\rm c}, \Phi_{\rm pr}, \Phi_\mu)$.

\subsubsection{Total mean expected number of events}

In the skymap of each event type, the angular distribution reflects the contributions of the fluxes of $\nu_e$, $\nu_\mu$, and $\nu_\tau$ (including their directional dependence), the effect on them of propagating inside the Earth~\cite{Arguelles:2021twb}, and the detector response to different flavors, implicitly contained in the IceCube HESE MC sample.

The total number of events of type $t$ in the $i$-th skymap pixel is the sum of the contributions from astrophysical neutrinos and atmospheric backgrounds, \ie,
\begin{equation}
 \label{equ:event_rate_total}
 N_i^t(\boldsymbol{\Theta})
 =
 \sum_{\alpha = e, \mu, \tau}
 N_{{\rm ast}, i}^{t,\alpha}(\boldsymbol{\theta})
 +
 N_{{\rm atm}, i}^t (\boldsymbol{\eta}) \;,
\end{equation}
where $\boldsymbol{\Theta} \equiv (\boldsymbol{\theta}, \boldsymbol{\eta})$.  This is the event rate that we use as the test hypothesis when computing the likelihood function.

\subsection{Modeling upcoming detectors with HESE MC}

The capabilities of the upcoming neutrino telescopes in Section~\ref{sec:future_detectors} to detect astrophysical neutrinos at HESE-like energies have not yet been modeled by their respective collaborations to the degree where we can account for their individual responses in detail. However, all but one of the detectors will use the same detection medium (water, solid or liquid), and rely on photomultipliers to detect and reconstruct their events. Therefore, we can approximate their responses by using the IceCube MC. We can estimate their event rates by scaling the weight of each event by the ratio of the detector's effective volume to IceCube's effective volume. To employ this in our analysis, we generated separate MC files for all the detectors that we simulate. TAMBO is the exception, but since its contribution to the collective event rates that we simulate is small, we treat it similar to the other detectors. For every detector except IceCube Gen2, we account for its different geographic position by adding to the true and simulated zenith and azimuthal angle of each event, the zenith and azimuthal coordinates of each of the new detectors. 

In this way, we account for the fact that detectors distributed all over the globe compensated for each other's field of view. Any detector not directly at the South Pole has a wider field of view, compared to IceCube and IC-Gen2, on the celestial sphere because it is aided by the Earth's rotation. Instantaneously, however, the Earth's volume always casts its shadow onto half the celestial sphere for each detector, and the fact that the direction of the resulting flux reduction changes over time does not affect the number of events we can expect as a result. However, for these detectors, their response to a neutrino flux from a particular source on the celestial sphere would be dependent on the time of day/year when the neutrinos from that source were detected. In our analysis, we are interested in the cumulative effects of combining several upcoming detectors, whose combined instantaneous field of view already covers the whole celestial sphere. We therefore do not model the effect of the Earth's rotation on each detector's individual field of view, instead estimating that is it time-independent. 

We simulate the astrophysical event rates, $N^{d,t}$, of morphology $t$, in an upcoming detector, $d$, by using Eq.~(\ref{equ:mc_reweigh_astro}). We draw the weight of each event from that detector's MC dataset, and replace the exposure time, $T$, with the exposure time of the simulated detector. 
Similarly, when simulating background contributions for each detector, the background event numbers in each pixel were extracted independently for each detector.

In the case of TAMBO, which would only be sensitive to $\tau$ neutrinos (and antineutrinos), we eliminated all events from its MC that were either caused by $e$ or $\mu$ neutrinos, or detected as anything other than double cascades.

\subsection{Modeling future constraint and discovery capabilities}

When forecasting constraints on $\boldsymbol{\Theta}$, we generate mock observed event samples using a prescription of the neutrino flux where the flavor composition is expanded in spherical harmonics (``fsh'' for ``flavor-only spherical harmonics'').  The (equal) flux of $\nu_\alpha$ or $\bar{\nu}_\alpha$ is
\begin{eqnarray}
 \label{equ:phi_new_harmonic}
 &&
 \Phi_{\nu_\alpha}^{\rm fsh}
 (E, \theta_z, \phi, \boldsymbol{\theta}^{\rm fsh})
 =
 \Phi_0
 \left(\frac{E}{100~{\rm TeV}}\right)^{-\gamma} 
 \times~
 \frac{1}{6} 
 \left[
 1
 +
 \sum_{\ell=1}^\infty \sum_{m = -\ell}^\ell 
 q^\alpha_{\ell, m}
 Y^m_\ell(\theta_z, \phi)
 \right] \;,
 \nonumber
\end{eqnarray}
where the flux parameters are $\boldsymbol{\theta}^{\rm fsh} \equiv (\Phi_0, \gamma, \{ q^\alpha_{\ell, m} \})$. 

Using this flux model, we generate the simulated observed event sample, $N_{{\rm obs}, i}^t$, as in \equ{event_rate_total}. When generating the astrophysical contribution in the observed event sample based on the above flux, the values of $q^\alpha_{\ell, m}$ are set either to zero---when placing upper limits on flavor anisotropy (\figu{cl_limits} below)---or to specific nonzero values---when forecasting its  measurement (Figs.~\ref{fig:all_ternaries_aniso_2040_atbfs}--\ref{fig:all_ternaries_aniso_2040_southpole} below).  When reporting results, we indicate the values of the nonzero coefficients when relevant. 

We used this flux model to test two questions: to which degree can we constrain directional flavor ratios if the true flux is flavor isotropic; and which degree of anisotropy is discoverable over time as more neutrino telescopes become operational? For the first question, in all instances, we simulated a purely isotropic, equal-flavor flux, \ie, all $\ell > 0$ harmonic moments were identically zero. The improvements made in our predictions can be seen in Fig.~\ref{fig:cl_limits}. We simulated time points for the years 2017 (IceCube HESE 7.5 year data release), 2025, 2030 and 2040.

For the second question, we chose to focus only on the azimuthally symmetric harmonic moments, \ie, $m=0$. We used the results of the constraints to make a first estimate of the discoverable flavor anisotropies for the dipole and the quadrupole. We made separate simulations for the dipole and the quadrupole. At each simulated time point, and for each harmonic moment, we calculated the constraint at a 90\% confidence level (C.L.). Realistically, how discoverable flavor anisotropies are would depend on the exact directional distribution that they manifest in nature. The azimuthally symmetric case we simulate, nevertheless, shows that the advent of future neutrino observatories could push the discovery potential below the current limits of anisotropy that we set, both for very large (dipole) and slightly smaller (quadrupole) angular sizes on the sky. The results can be seen in Fig.~\ref{fig:cl_disco}. 

Lastly, for illustration purposes, we generate three additional simulations. One replicates the current best-fit harmonic moments from the HESE 7.5 year data, and showed how discoverable this scenario would be in 2040, with all planned detectors in operation. The simulated harmonic moments for this case can be found in Table~\ref{tab:harmonic_moment_data}, and the recovered posteriors for the pixels are shown in Fig.~\ref{fig:all_ternaries_aniso_2040_atbfs}. We also simulate a ``large'' anisotropy, with $q^e_{1,0}=1.5, q^\mu_{1,0}=-0.5, q^\tau_{1,0} = -1$ and $q^e_{1,1} = 0.2i, q^\mu_{1,1} = -1.2i, q^\tau_{1,1} = -i$, to demonstrate the gain in sensitivity from geographically-spreading the proposed detectors. We simulate event rate skymaps from the same underlying flux parameters for the year 2040 in the case that (a) all detectors were at their proposed locations and (b) that all detectors are artificially at the South Pole. To quantify the improvement in sensitivity from case (b) to case (a), we calculate the Bayes factor using the reference flux in \equ{phi_ref}. {The null hypothesis used in determining the Bayes factor was a statistical fit with the reference model in Eq.~\ref{equ:phi_ref}.}

\section{Statistical methods}

We created 12 directional bins on the celestial sphere using HEALpix (parameter $n_\text{side}=1$, in the RING ordering). The events were binned in the resulting 12 pixels tessellating the sky according to their best-fit reconstructed direction. Each pixel is about 90$^\circ$ wide: given the paucity of present data, this coarse tessellation prevents empty pixels, but renders structures smaller than tens of degrees in the sky unresolvable.

The flux in \equ{flav_flux_earth} is defined so that integrating it over the sphere, and setting $f_{\alpha, \oplus} (\theta_z, \phi) = 1/3$, recovers the same shape as the reference power-law flux, \equ{phi_ref}, thus allowing for meaningful comparison between them.

We model detector, $d = \text{IC}$, by the native IceCube HESE MC, modified to include real and reconstructed azimuthal event coordinates, as outlined in Section~\ref{sec:mc_modifications}. We include more detectors when we simulate different astrophysical flavor flux scenarios for the future time points. The model of the astrophysical flux used in this analysis is the same as in \equ{flav_flux_earth}, the background event rates are given by Eq.~(\ref{eq:background_event_rates}), and the likelihood is given in Eq.~(\ref{equ:likelihood_pixel}). 

Table~\ref{tab:fit_params} collects all our model parameters, $\boldsymbol{\Theta}$, and their priors, $\pi(\boldsymbol{\Theta})$. In total, we fit a 29-dimensional parameter space. In all our fits to data, both real and simulated, we use the same set of parameters and priors as in Table~\ref{tab:fit_params} to obtain our results.

\begin{table*}[h!]
 \caption{\label{tab:fit_params}\textbf{\textit{Free model parameters and their priors from a fit to the IceCube 7.5-year HESE event sample.}} The three-dimensional Dirichlet distributions used as priors for the flavor-composition parameters ensure that $f_{e, i} + f_{\mu, i} + f_{\tau, i} = 1$ in each pixel.}\vspace{10pt}
  \centering
  \renewcommand{\arraystretch}{1.3}
  \begin{tabular}{cccc}
  \hline\hline
  \multicolumn{3}{c}{Parameter} &
  \multirow{2}{*}{Prior} \\
  \cline{1-3}
  \multirow{1}{*}{Name} &
  \multirow{1}{*}{Units} &
  \multirow{1}{*}{Description} &
  \\
  \hline
  \multicolumn{4}{c}{Flux parameters} \\
  \hline \vspace{0.3em}
  \multirow{2}{*}{$\Phi_{0}$} &
  \multirow{2}{*}{$10^{-18}$~GeV$^{-1}$~cm$^{-2}$~s$^{-1}$~sr$^{-1}$} &
  \multirow{2}{*}{\makecell{All-flavor\\flux norm.~at 100~TeV}} &
  \multirow{2}{*}{\makecell{Normal,\\$\mu = 5.68$, $\sigma = 1.2$}} \\[0.8em] 
  \multirow{2}{*}{$\gamma$} &
  \multirow{2}{*}{$\cdots$} &
  \multirow{2}{*}{Spectral index} &
  \multirow{2}{*}{\makecell{Normal,\\$\mu = 2.89$, $\sigma = 0.23$}} \\[0.8em] 
  \multirow{2}{*}{$f_{e,i}$, $f_{\mu,i}$} &
  \multirow{2}{*}{$\cdots$} &
  \multirow{2}{*}{\makecell{Flavor composition,\\pixel $i = 1,\cdots,12$}} &
  \multirow{2}{*}{3D Dirichlet} \\\\
  \hline
  \multicolumn{4}{c}{Nuisance parameters} \\
  \hline
  \multirow{2}{*}{$\Phi_{\text{c}}$} &
  \multirow{2}{*}{$\cdots$} &
  \multirow{2}{*}{Flux norm., convent.~atm.~$\nu$} &
  \multirow{2}{*}{\makecell{Normal,\\ $\mu=1.0$, $\sigma=0.4$}} \\
  \\
  $\Phi_{\text{pr}}$ &
  $\cdots$ &
  Flux norm., prompt~atm.~$\nu$ &
  Uniform $\in \left[0, 10\right]$ \\
  \multirow{2}{*}{$\Phi_{\mu}$} &
  \multirow{2}{*}{$\cdots$} &
  \multirow{2}{*}{Flux norm., atm.~$\mu$} &
  \multirow{2}{*}{\makecell{Normal,\\$\mu=1.0$, $\sigma=0.5$}} \\\\\hline\hline
  \end{tabular}
\end{table*}


\subsection{Likelihood}

For one detector, $d$, in the $i$-th skymap pixel, the Poisson likelihood function for events of type $t$ is
\begin{equation}
 \label{equ:likelihood_pixel}
 \mathcal{L}_i^{d,t}(\boldsymbol{\Theta})
 =
 \frac{ N_i^{d,t}(\boldsymbol{\Theta})^{N_{{\rm obs}, i}^{d,t}} 
 e^{-N_i^{d,t}(\boldsymbol{\Theta})} } 
 { N_{{\rm obs}, i}^{d,t}! } \;,
\end{equation}
where $N_i^{d,t}(\boldsymbol{\Theta})$ is the mean expected number of events in detector $d$, for a certain test value of $\boldsymbol{\Theta}$, given by \equ{event_rate_total}, and $N_{{\rm obs}, i}^{d,t}$ is the number of observed events in the $i$-th pixel in detector $d$, either in real data, \ie, the IceCube 7.5-year HESE sample, or in  mock data.  The all-sky likelihood for event type $t$ is the product of all pixels, \ie,
\begin{equation}
 \mathcal{L}^{d,t}(\boldsymbol{\Theta})
 =
 \prod_{i=1}^{N_{\rm pix}}
 \mathcal{L}_i^{d,t}(\boldsymbol{\Theta}) \;,
\end{equation}
and the total likelihood includes all event types, $t$, in all detectors, $d$, \ie,
\begin{equation}
 \mathcal{L}(\boldsymbol{\Theta})
 =\prod_d
 \mathcal{L}^{d,\rm c}(\boldsymbol{\Theta})
 \mathcal{L}^{d,\rm tr}(\boldsymbol{\Theta})
 \mathcal{L}^{d,\rm dc}(\boldsymbol{\Theta}) \;.
\end{equation}
The joint posterior distribution is
\begin{equation}
 \label{equ:posterior}
 \mathcal{P}(\boldsymbol{\Theta}) 
 =
 \mathcal{L}(\boldsymbol{\Theta})
 \pi(\boldsymbol{\Theta}) \;.
\end{equation}

\subsubsection{Recovering spherical harmonic moments from flavor composition}
\label{sec:anafast}
After recovering the joint posterior of all the fit parameters, we isolate the part of the posterior parametrization the flavor ratios in each pixel. This yields three sets of skymaps, one for each neutrino flavor. Each skymap posterior of each flavor now contains the flavor ratios that were sampled together, capturing the possible correlation between the flavor composition recovered in adjacent pixels. In this way, we re-cast the pixel-binned flavor composition posteriors as flavor skymap posteriors.  For each skymap of each flavor in the posterior, we use the native HEALpix \texttt{anafast} method to obtain the first three spherical harmonic moments from an $n_\text{side} = 1$ map, \ie,  the monopole, dipole and quadrupole moments. In this way, we transform the joint posteriors on the pixels to joint posteriors on the flavor-specific harmonic moments.

\subsection{Simulating future constraint and discovery possibilities}
\label{sec:stats_simulations}
To infer directional flavor composition at present, we used the HESE dataset released by the IceCube collaboration, consisting of 102 events, distributed across the sky. 

In Figure~\ref{fig:future-detectors} and Table~\ref{tab:detectors}, we show the upcoming detectors that we simulate, their global positions, effective volumes and year of starting (ending) operation. To quantify how much the analysis we performed on the current HESE data can improve with the advent of more data, and from detectors distributed globally, we simulated several timesteps, corresponding to the expected startdates of upcoming detectors; 2017 (HESE release year), 2025 (IceCube only, more data), 2030 (IceCube, KM3NeT and Baikal-GVD), and 2040 (IceCube + IceCube Gen2, KM3NeT, Baikal-GVD, P-One and TAMBO). For each simulation, we kept the astrophysical all-flavor flux parameters fixed at their best-fit values as reported by IceCube, \ie, $\Phi_0 = 5.68\cdot 10^{-18}$ GeV$^{-1}$cm$^{-2}$s$^{-1}$sr$^{-1}$ and $\gamma = 2.89$. We also kept the background flux normalization parameters fixed at $\Phi_\text{c} =1$, $\Phi_\text{pr}=0$ and $\Phi_\mu = 1$. 

To test to which degree we can constrain a truly isotropic flavor flux, we simulated, at each time step, a flavor-isotropic flux, \ie, $q^\alpha_{\ell>0, m} =0$ for all flavors $\alpha = e,\mu,\tau$. Our fits to the HESE data recover non-zero best fits for flavor anisotropies, that are nevertheless compatible with isotropy at $1\sigma$. We used those best-fit values to test how discoverable such an anisotropy would be with the advent of more detectors. The values we simulate are the best-fit spherical harmonic values, listed in Table~\ref{tab:harmonic_moment_data}.  

The degree of flavor anisotropy discoverable at each time point was estimated using the spherical harmonic power spectrum for each neutrino flavor. The power spectrum coefficients for each flavor are defined for each harmonic $\ell$ with the identity
\begin{equation}
    C_\ell^\alpha = \frac{2\pi}{\ell+1}\sum_{m=-\ell}^\ell \left|q^\alpha_{\ell,m}\right|^2\, .
\end{equation}

Due to the polar angle bias in the field of view of IceCube, unless otherwise indicated, we made the pessimistic choice to always simulate anisotropies using the azimuthally asymmetric and polar-asymmetric harmonic moments ($m=0$). This is a pessimistic choice because IceCube's polar-asymmetric field of view leads to higher uncertainties in directional flavor reconstruction along the polar axis. For later time points, when more detectors are included, this assumption becomes less pessimistic, as the global distribution of the telescopes compensates for the obstructed field of view of one detector alone. 

\subsection{Measuring the anisotropies in the all-flavor flux}

Our flavor flux model still implicitly assumes that the all-flavor high-energy astrophysical neutrino flux is isotropic. We also want to quantify the degree to which this is the case in present data, and to which degree this can be constrained/discovered in the future. To do that, we performed additional, and independent, analyses, using a flux prescription that allows for directional dependence in the all-flavor flux (``dda'', for ``directionally dependent all-flavor flux''), but assumes isotropy in the flavor composition, \ie,
\begin{eqnarray}
 \label{equ:tot_aniso_evt_rate}
 \Phi_{\nu_\alpha}^{\rm dda}
 (E, \theta_z, \phi, \boldsymbol{\theta})
 &=&
 \frac{1}{N_\text{pix}}
 \frac{f_{\alpha, \oplus}}{2} 
 \left(\frac{E}{100~{\rm TeV}}\right)^{-\gamma} 
 \nonumber \times~
 \Phi_{\rm all}(\theta_z, \phi) \;,
\end{eqnarray}
where $f_{\alpha, \oplus}$ is the all-sky flavor composition, and the all-flavor flux $\Phi_{\rm all} (\theta_z, \phi) = \Phi_{0, k}$, with $k$ being the pixel that contains the direction $(\theta_z, \phi)$. The flux parameters are $\boldsymbol{\theta}^{\rm dda} \equiv (f_{e,\oplus}, f_{\mu,\oplus}, \gamma, \{ \Phi_{0,i} \}_{i=1}^{N_{\rm pix}})$, with $N_{\rm pix} = 12$.  (This expression is the complement of \equ{flav_flux_earth}, which allows for directional dependence in the flavor composition, but assumes isotropy in the all-flavor flux.)

To compute the event rates, we used the same MC reweighing technique as for the directional flavor analysis, and the same likelihood prescription. We choose similar priors as for our flavor anisotropy analysis: for $f_{e, \oplus}$ and $f_{\mu, \oplus}$, we use a three-dimensional Dirichlet distribution and, for $\gamma$, a Gaussian; see Table~\ref{tab:fit_params}.  For each $\Phi_{0, k}$, we use a uniform distribution in $[0,20]$, in units of $10^{-18}$~GeV$^{-1}$~cm$^{-2}$~s$^{-1}$~sr$^{-1}$, which grants the fit considerable freedom to spot anisotropies.  From the resulting posteriors of $\Phi_{0, k}$, we used the HEALPix routine {\tt anafast} to extract the posteriors of the coefficients of the spherical-harmonic expansion of the all-flavor flux and, from them, we compute the posteriors of the power in the dipole and quadrupole moments, $C_{\ell = 1}^{\rm all}$ and $C_{\ell = 2}^{\rm all}$.

We use this model to test whether our flavor flux model in \equ{flav_flux_earth} recovers valid results if our assumption of an all-flavor flux isotropy is broken. See Appendix~\ref{app:method_validation} for details.


\section{Results}

\subsection{HESE directional flavor composition}

 \begin{figure*}[h!]
  \centering
  \includegraphics[width=1.05\textwidth]{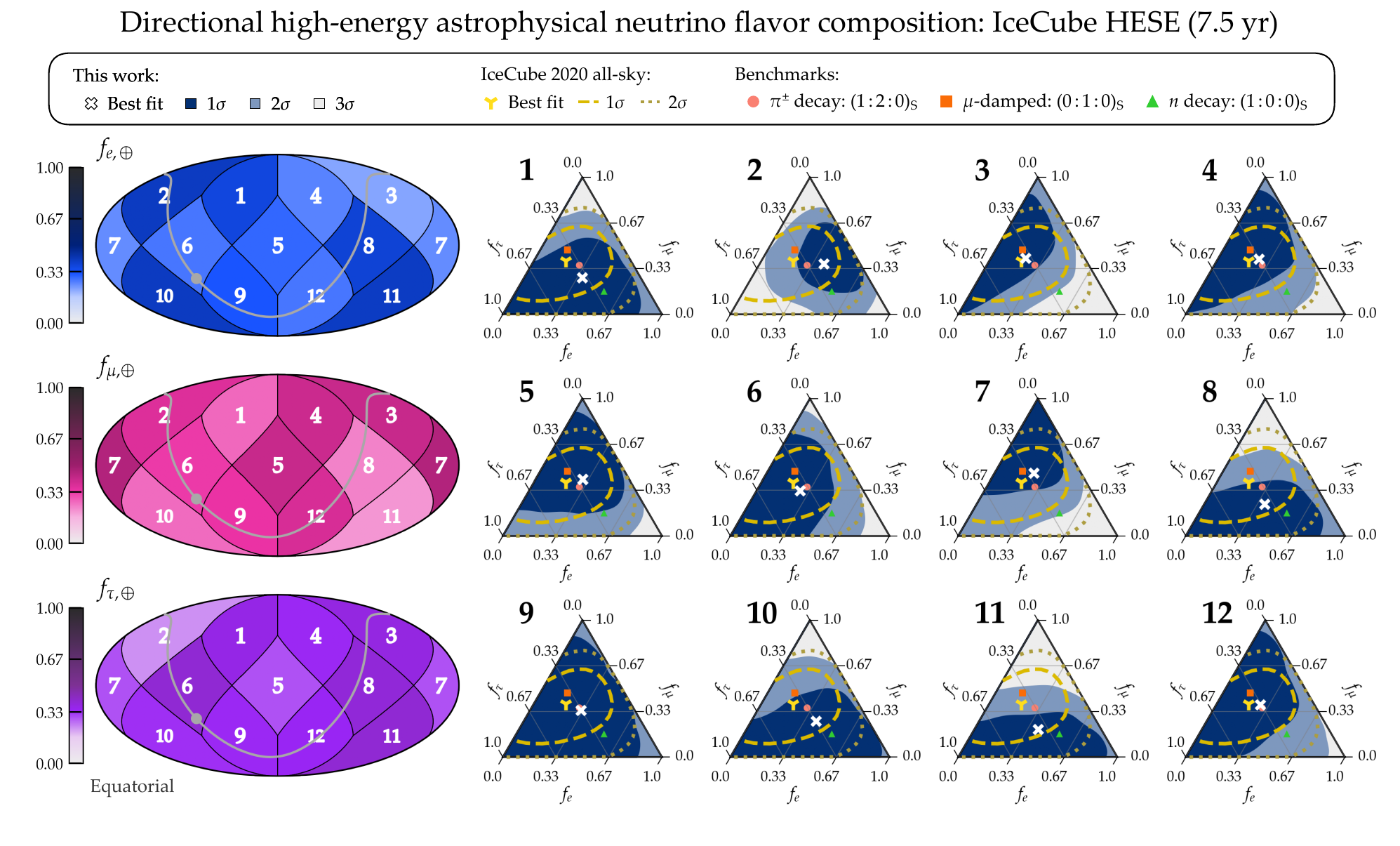}\vspace{-20pt}
  \caption{\textbf{\textit{Directional flavor composition of high-energy astrophysical neutrinos in present-day IceCube data.}}  Results are obtained using the public 7.5-year IceCube HESE sample~\cite{IceCube:2020wum}.  The skymap tessellation is from HEALPix (with $N_{\rm side} = 1$); pixel numbering follows the ring ordering~\cite{Gorski:2004by, healpix_url}.  \textit{The results are compatible with isotropy in all flavors.}\\}
  \label{fig:all_ternaries_data}
 \end{figure*}

The directional flavor measurements of the current HESE data, as well as the allowed anisotropic spherical harmonic parameters were derived with a fit to the 102 events in the HESE data release, according to the statistical analysis we outlined above. In our analysis, we find that the flavor-composition posteriors overlap in all pixels at $1\sigma$, and that the recovered harmonic moments are compatible with flavor isotropy for all flavors. 

\begin{table*}[h!]
 \caption{\label{tab:fit_params_combined}\textbf{\textit{Free model parameters, best-fit values, and allowed ranges from a fit to the IceCube 7.5-year HESE event sample.}} Allowed parameter ranges are 68\% one-dimensional marginalized credible intervals. The three-dimensional Dirichlet distributions ensure that $f_{e,i} + f_{\mu,i} + f_{\tau,i} = 1$ in each pixel.}\vspace{10pt}
  \centering
  \renewcommand{\arraystretch}{1.3}
  \begin{tabular}{cccccc}
  \hline\hline
  \multicolumn{2}{c}{Parameter} &
  \multicolumn{4}{c}{Fit to 7.5-yr IceCube HESE sample} \\
  \cline{1-2}
  \cline{3-6}
  \multirow{2}{*}{Name} &
  \multirow{2}{*}{Units} &
  \multirow{2}{*}{all-flavor} &
  \multirow{2}{*}{$f_{e,i}$} &
  \multirow{2}{*}{$f_{\mu,i}$} &
  \multirow{2}{*}{\makecell{$f_{\tau,i}$\\$= 1-f_{e,i}-f_{\mu,i}$}} \\
  \\
  \hline
  \multicolumn{6}{c}{Flux parameters} \\
  \hline 
  $\Phi_{0}$ &
  $10^{-18}$~GeV$^{-1}$~cm$^{-2}$~s$^{-1}$~sr$^{-1}$ &
  $5.24^{+0.79}_{-0.73}$ &
  $\cdots$ &
  $\cdots$ &
  $\cdots$ \\[0.2em] 
  $\gamma$ &
  $\cdots$ &
  $2.91 \pm 0.19$ &
  $\cdots$ &
  $\cdots$ &
  $\cdots$ \\[0.2em] 
  $f_{e,1}$, $f_{\mu,1}$ &
  $\cdots$ &
  $\cdots$ &
  $0.34^{+0.27}_{-0.22}$ &
  $0.23^{+0.25}_{-0.16}$ &
  $0.34^{+0.29}_{-0.23}$ \\[0.2em] 
  $f_{e,2}$, $f_{\mu,2}$ &
  $\cdots$ &
  $\cdots$ &
  $0.40^{+0.21}_{-0.21}$ &
  $0.36^{+0.21}_{-0.19}$ &
  $0.18^{+0.24}_{-0.13}$ \\[0.2em] 
  $f_{e,3}$, $f_{\mu,3}$ &
  $\cdots$ &
  $\cdots$ &
  $0.18^{+0.22}_{-0.13}$ &
  $0.40^{+0.25}_{-0.24}$ &
  $0.35^{+0.28}_{-0.22}$ \\[0.2em] 
  $f_{e,4}$, $f_{\mu,4}$ &
  $\cdots$ &
  $\cdots$ &
  $0.22^{+0.25}_{-0.16}$ &
  $0.39^{+0.28}_{-0.25}$ &
  $0.30^{+0.31}_{-0.21}$ \\[0.2em] 
  $f_{e,5}$, $f_{\mu,5}$ &
  $\cdots$ &
  $\cdots$ &
  $0.26^{+0.25}_{-0.18}$ &
  $0.40^{+0.25}_{-0.24}$ &
  $0.25^{+0.21}_{-0.18}$ \\[0.2em] 
  $f_{e,6}$, $f_{\mu,6}$ &
  $\cdots$ &
  $\cdots$ &
  $0.23^{+0.26}_{-0.16}$ &
  $0.30^{+0.27}_{-0.20}$ &
  $0.38^{+0.29}_{-0.25}$ \\[0.2em] 
  $f_{e,7}$, $f_{\mu,7}$ &
  $\cdots$ &
  $\cdots$ &
  $0.21^{+0.24}_{-0.15}$ &
  $0.46^{+0.27}_{-0.27}$ &
  $0.25^{+0.30}_{-0.18}$ \\[0.2em] 
  $f_{e,8}$, $f_{\mu,8}$ &
  $\cdots$ &
  $\cdots$ &
  $0.36^{+0.27}_{-0.24}$ &
  $0.20^{+0.22}_{-0.14}$ &
  $0.27^{+0.28}_{-0.25}$ \\[0.2em] 
  $f_{e,9}$, $f_{\mu,9}$ &
  $\cdots$ &
  $\cdots$ &
  $0.28^{+0.28}_{-0.20}$ &
  $0.32^{+0.28}_{-0.22}$ &
  $0.30^{+0.31}_{-0.21}$ \\[0.2em] 
  $f_{e,10}$, $f_{\mu,10}$ &
  $\cdots$ &
  $\cdots$ &
  $0.39^{+0.28}_{-0.25}$ &
  $0.22^{+0.25}_{-0.15}$ &
  $0.29^{+0.30}_{-0.20}$ \\[0.2em] 
  $f_{e,11}$, $f_{\mu,11}$ &
  $\cdots$ &
  $\cdots$ &
  $0.38^{+0.29}_{-0.26}$ &
  $0.17^{+0.20}_{-0.12}$ &
  $0.37^{+0.31}_{-0.25}$ \\[0.2em] 
  $f_{e,12}$, $f_{\mu,12}$ &
  $\cdots$ &
  $\cdots$ &
  $0.23^{+0.27}_{-0.16}$ &
  $0.36^{+0.29}_{-0.25}$ &
  $0.30^{+0.31}_{-0.21}$ \\[0.2em] 
  \hline
  \multicolumn{6}{c}{Nuisance parameters} \\
  \hline
  $\Phi_{\text{c}}$ &
  $\cdots$ &
  $0.98^{+0.38}_{-0.39}$ &
  $\cdots$ &
  $\cdots$ \\[0.2em] 
  $\Phi_{\text{pr}}$ &
  $\cdots$ &
  $0.13^{+0.21}_{-0.10}$ &
  $\cdots$ &
  $\cdots$ &
  $\cdots$ \\[0.2em] 
  $\Phi_{\mu}$ &
  $\cdots$ &
  $0.99^{+0.32}_{-0.30}$ &
  $\cdots$ &
  $\cdots$ &
  $\cdots$ \\\hline\hline
  \end{tabular}
\end{table*}

\begin{table*}[h!]
 \caption{\label{tab:harmonic_moment_data}\textbf{\textit{Present-day best-fit values and allowed ranges of the coefficients of the flavor spherical-harmonics expansion, $\boldsymbol{q_{\ell, m}^\alpha}$, and power spectrum, $\boldsymbol{C_\ell^\alpha}$.}}  Allowed parameter ranges are 68\% one-dimensional marginalized credible intervals, computed by combining the posteriors of the directional flavor composition, $f_{e,i}$ and $f_{\mu,i}$, themselves derived from a fit to the IceCube 7.5-year HESE sample.}\vspace{10pt}
  \centering
  \renewcommand{\arraystretch}{1.3}
  \begin{tabular}{ccccc}
  \hline\hline
   Parameter & 
   $\nu_e$ sky ($\alpha = e$) & 
   $\nu_\mu$ sky ($\alpha = \mu$) & 
   $\nu_\tau$ sky ($\alpha = \tau$) &
   All-flavor sky \\ 
   \hline
   \multicolumn{5}{c}{Dipole, $\ell=1$} \\
   \hline
   $q_{1,0}^\alpha$ &
   $-0.14\pm 0.39$ &
   $0.22\pm0.36$ &
   $-0.08\pm0.39$ &
   $-0.2\pm0.26$ \\[0.2em] 
   ${\rm Re} (q_{1,1}^\alpha)$ &
   $0.07\pm 0.30$ &
   $-0.04\pm 0.30$ &
   $-0.03\pm0.32$ &
   $0.13\pm0.22$ \\ [0.2em] 
   ${\rm Im} (q_{1,1}^\alpha)$ &
   $0.09\pm0.31$ &
   $0.00\pm0.28$ &
   $-0.09\pm0.32$ &
   $0.03\pm0.21$ \\[0.2em] 
   $C_1^\alpha$&
   $0.05^{+0.14}_{-0.05}$ &
   $0.06^{+0.12}_{-0.06}$&
   $0.05^{+0.12}_{-0.05}$&
   $0.03^{+0.09}_{-0.03}$\\
   \hline
   \multicolumn{5}{c}{Quadrupole, $\ell=2$} \\
   \hline
   $q_{2,0}^\alpha$&
   $0.17\pm0.63$&
   $-0.23\pm0.61$&
   $-0.01\pm0.66$&
   $-0.29\pm0.42$\\[0.2em] 
   ${\rm Re} (q_{2,1}^\alpha)$ &
   $-0.15\pm0.31$&
   $0.25\pm0.29$&
   $-0.10\pm0.31$&
   $0.04\pm0.21$\\[0.2em] 
   ${\rm Im} (q_{2,1}^\alpha)$ &
   $0.18\pm0.31$ &
   $-0.13\pm0.28$&
   $-0.03\pm0.31$&
   $0.03\pm0.20$\\[0.2em] 
   ${\rm Re} (q_{2,2}^\alpha)$ &
   $-0.11\pm0.29$ &
   $0.29\pm 0.29$&
   $-0.19\pm0.32$&
   $-0.02\pm0.20$\\[0.2em] 
   ${\rm Im} (q_{2,2}^\alpha)$ &
   $0.04\pm0.39$ &
   $0.17\pm0.37$&
   $-0.20\pm0.40$&
   $0.05\pm0.27$\\[0.2em] 
   $C_2^\alpha$&
   $0.08^{+0.21}_{-0.08}$ &
   $0.12^{+0.24}_{-0.12}$ &
   $0.11^{+0.19}_{-0.11}$ &
   $0.05^{+0.10}_{-0.05}$\\\hline\hline
  \end{tabular}
\end{table*}

Figure~\ref{fig:all_ternaries_data} shows the present-day measurement of the directional flavor composition, based on the 7.5-year IceCube HESE sample \cite{IC75yrHESEPublicDataRelease}.  The results are compatible with isotropy (Bayes factor $\mathcal{B} = 0.9 \pm 0.4$). The skymaps in the figure show the best-fit flavor composition for each flavor and each pixel. The best-fit skymaps show some degree of anisotropy, but the allowed regions for the flavor composition in each individual pixel (shown in the corresponding ternary plots) all have overlapping $1\sigma$ contours. In the figure, we included the contours of IceCube's own sky-averaged flavor composition analysis of the HESE data. In data-sparse pixels, especially pixel 9, the allowed regions are far wider than in the more populated pixels. Our results for the HESE 7.5-year analysis should then be taken as indicative; and a more systematized analysis should be carried out with more data in the near-future. Overall, all pixels also overlap significantly with the all-sky flavor analysis contours from IceCube. 

Table~\ref{tab:fit_params_combined} summarizes the one-dimensional allowed regions for all the flavor composition in each pixel, as well as the all-flavor parameters used in our fit. For the astrophysical flux, $\Phi_0$ and $\gamma$ are both compatible with the fits performed by the IceCube Collaboration for the flux normalization and spectral index \cite{IceCube:2020wum}. The background normalization parameters, $\Phi_\text{c}, \Phi_\text{pr}$, and $\Phi_\mu$ are all also compatible with IceCube's measurements. For the flavor ratio parameters, capturing flavor anisotropies, we find that all pixels overall are compatible with flavor isotropy, as well as the flavor splitting resulting from a pion decay production mechanism at sources. The one-dimensional allowed regions presented in this table do not capture the correlation between the flavor ratios in each pixel and are given here only as indicative.. 

Table~\ref{tab:harmonic_moment_data} converts the posteriors of the pixelized flavor composition maps to posteriors of the spherical harmonics---the dipole and quadrupole--that can be inferred from our our tessellated skymaps.

\begin{figure}[h!]
 \centering
 \includegraphics[width=0.49\columnwidth]{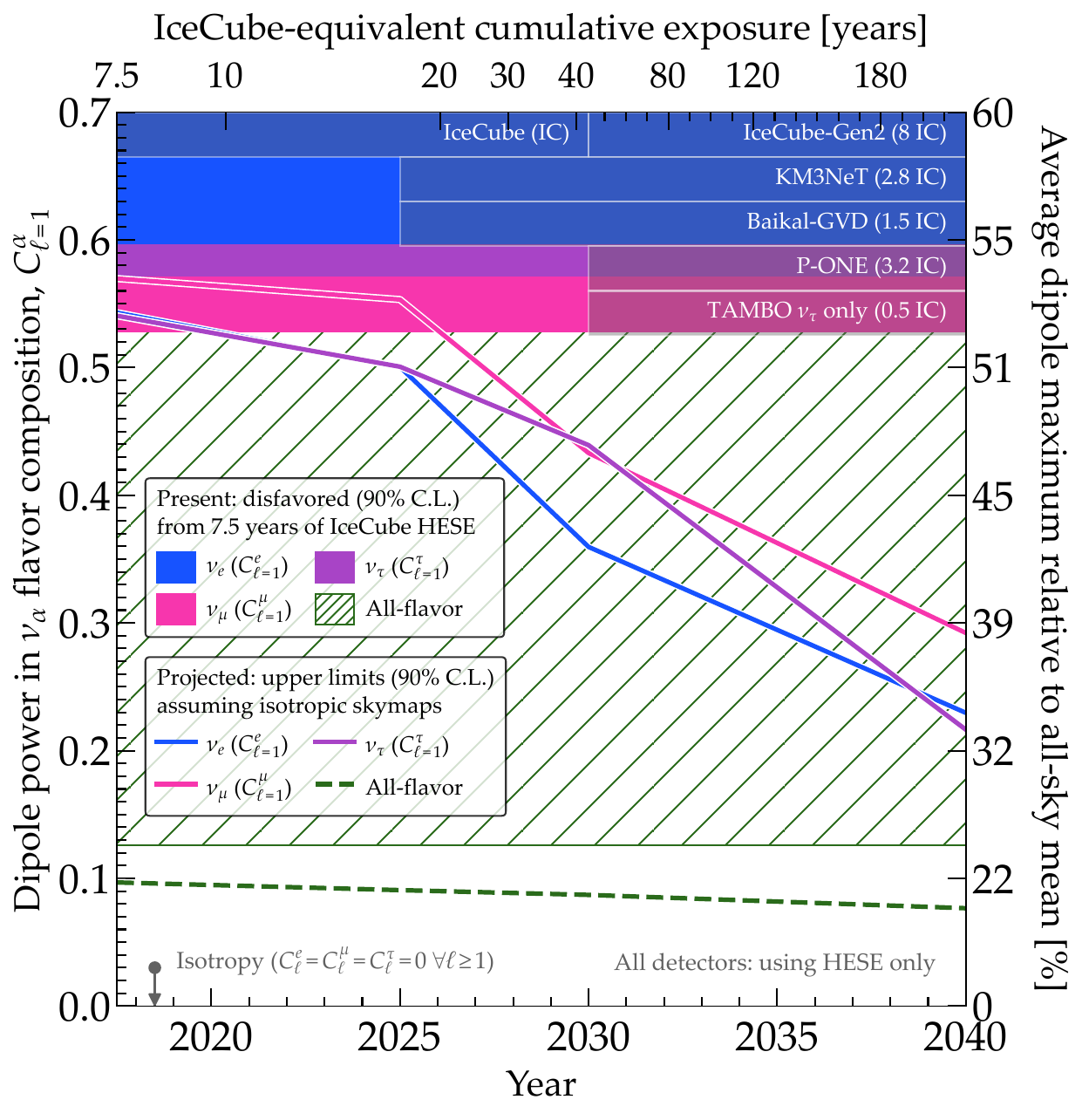}\hspace{0pt}
 \includegraphics[width=0.49\columnwidth]{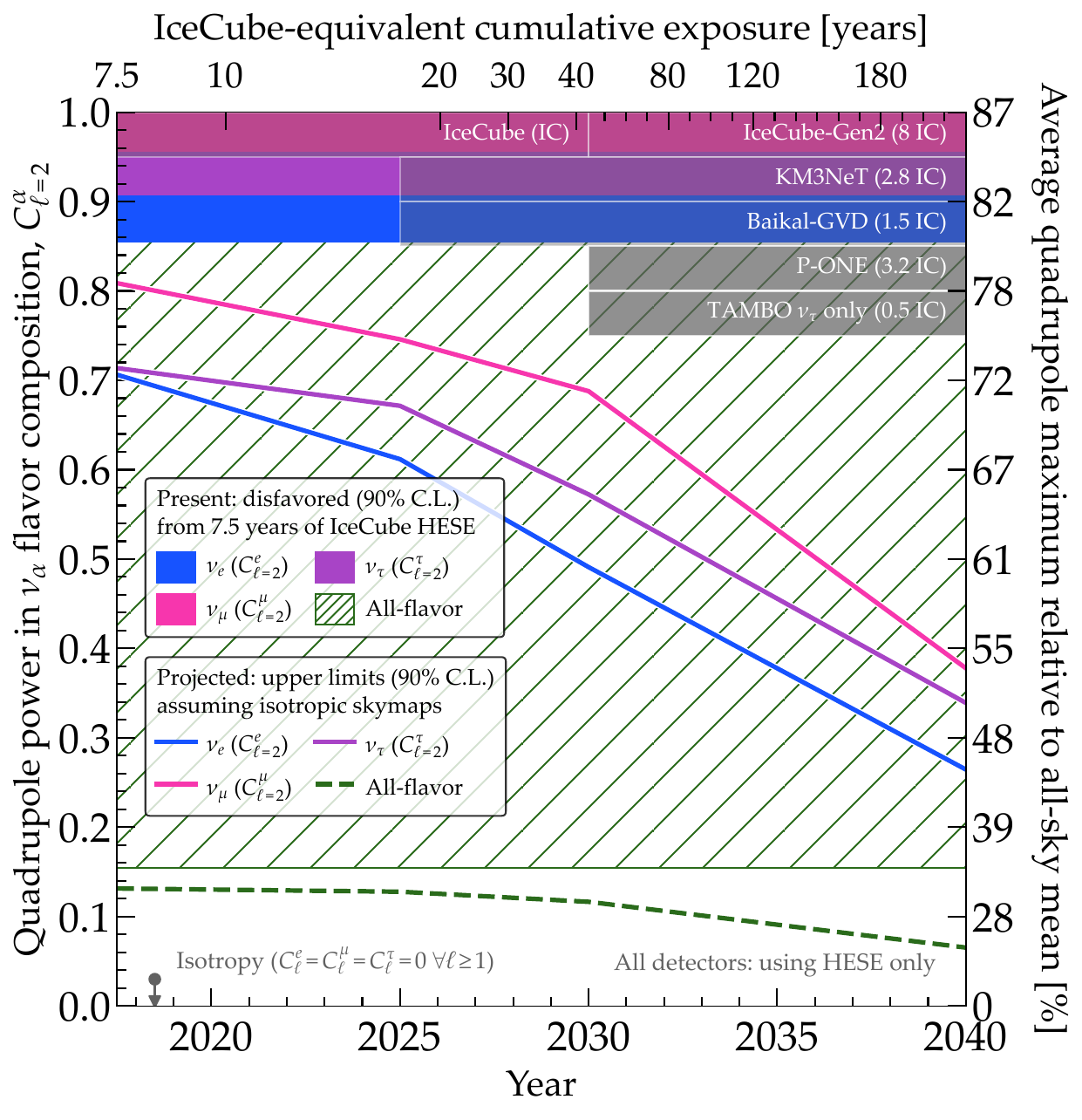}
 \vspace{-0pt}
 \caption{\textbf{\textit{Upper limits on the dipole power (left) and quadrupole power (right) in the $\nu_\alpha$ flavor-ratio skymap, $\boldsymbol{C_{\ell}^\alpha}$.}} 
 Projected limits are derived assuming that the skymaps of $\nu_e$, $\nu_\mu$ and $\nu_\tau$ are isotropic, and using the combined detection of upcoming neutrino telescopes, whose sizes relative to IceCube and tentative start dates are indicated as insets.  Table~\ref{tab:harmonic_moment_data} contains the numerical values of the present-day upper limits. 
 }\vspace{20pt}
 \label{fig:cl_limits}
\end{figure}

\begin{figure}[h!]
 \centering
 \includegraphics[width=0.49\columnwidth]{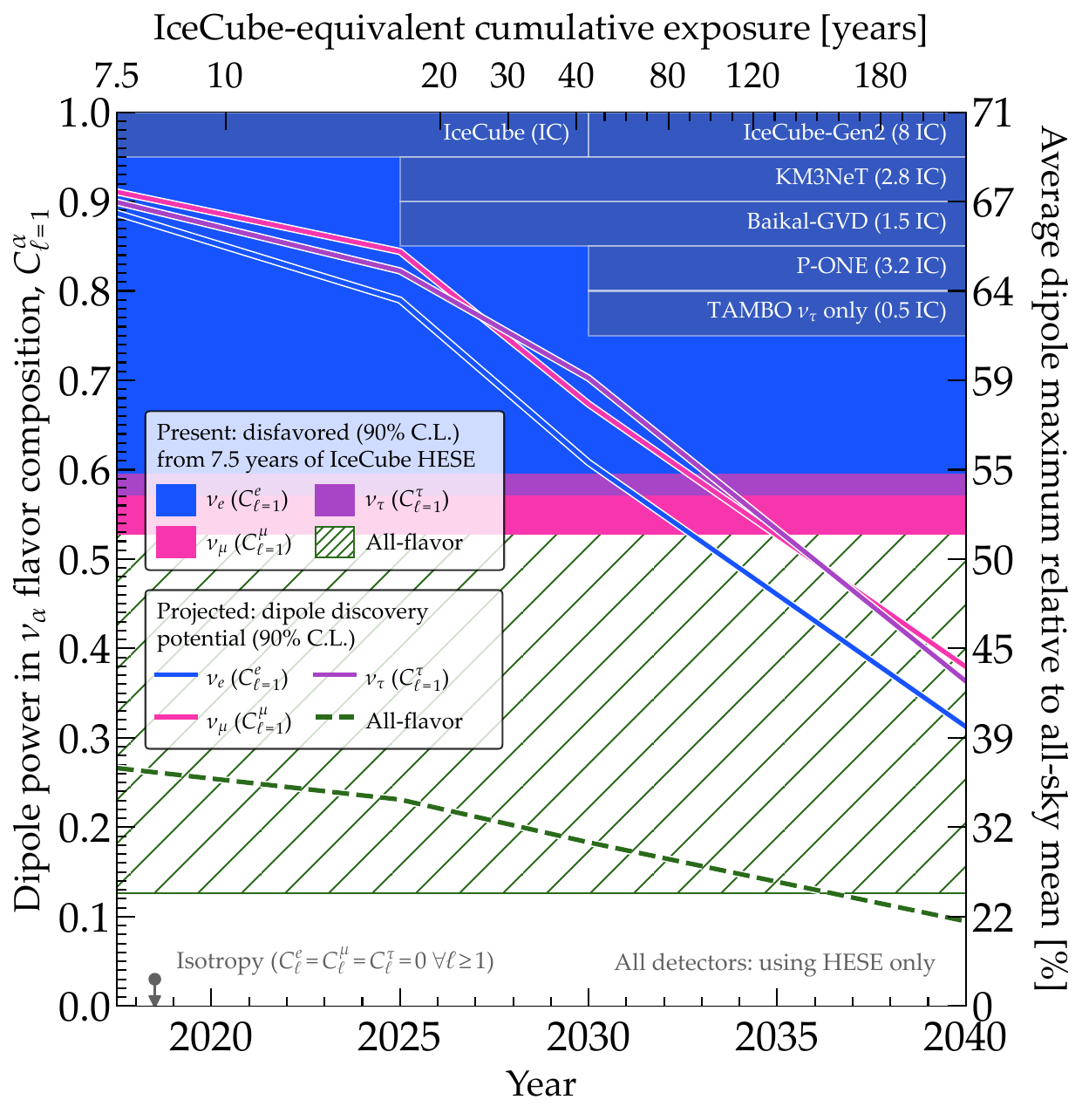}\hspace{0pt}
 \includegraphics[width=0.49\columnwidth]{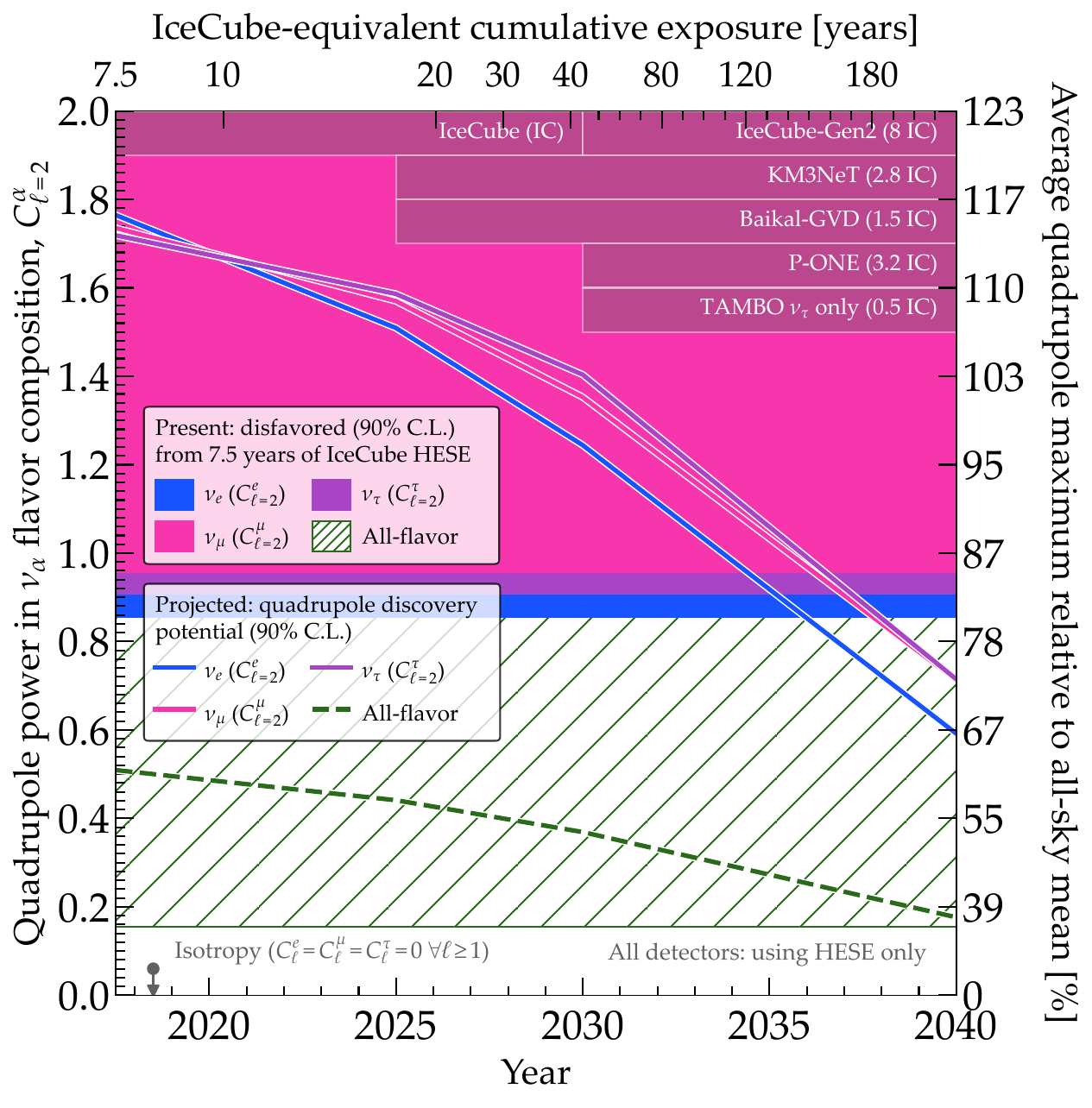}
 \vspace{-0pt}
 \caption{\textbf{\textit{Minimum discoverable dipole power (left) and quadrupole power (right) in the $\boldsymbol{\nu_\alpha}$ flavor-composition and all-flavor skymaps, $\boldsymbol{C_{\ell}^\alpha}$ and $\boldsymbol{C_{\ell}^{\rm all}}$.}} The present-day limits are the same as in \figu{cl_limits}.  The projected discovery potential for the flavor coefficients is derived by varying only $q_{1,0}^e$, $q_{1,0}^\mu$, and $q_{1,0}^\tau$ for the dipole (left), and only $q_{2,0}^e$, $q_{2,0}^\mu$, and $q_{2,0}^\tau$ for the quadrupole (right). }
 \label{fig:cl_disco}
\end{figure}

\subsection{Future improvements}

\subsubsection{Forecasting isotropy constraints}

Figure~\ref{fig:cl_limits} shows the upper limits on the dipole and quadrupole moments of the flavor power spectrum, $C_{\ell = 1}^\alpha$ and $C_{\ell = 2}^\alpha$. The present-day limits are extracted from the posteriors in \figu{all_ternaries_data}.  The projected limits are extracted from mock future posteriors of $C_{\ell}^\alpha$, generated assuming isotropy in the skymaps of all flavors individually, and, independently for the all-flavor anisotropy limits, assuming the all-flavor sky is also isotropic. Notably, the degree to which we can constrain flavor isotropy improves by up to 40\% by 2040 for the dipole, and by up to 50\% for the quadrupole, if all the detectors we consider are in operation. For both the dipole and quadrupole, the electron and tau anisotropies improve slightly faster with the advent of more data and more detectors.

\subsubsection{Forecasting anisotropy discovery}

 \begin{figure*}[t!]
  \centering
  \includegraphics[trim={0 0.3cm 0 0.25cm}, clip, width=1\columnwidth]{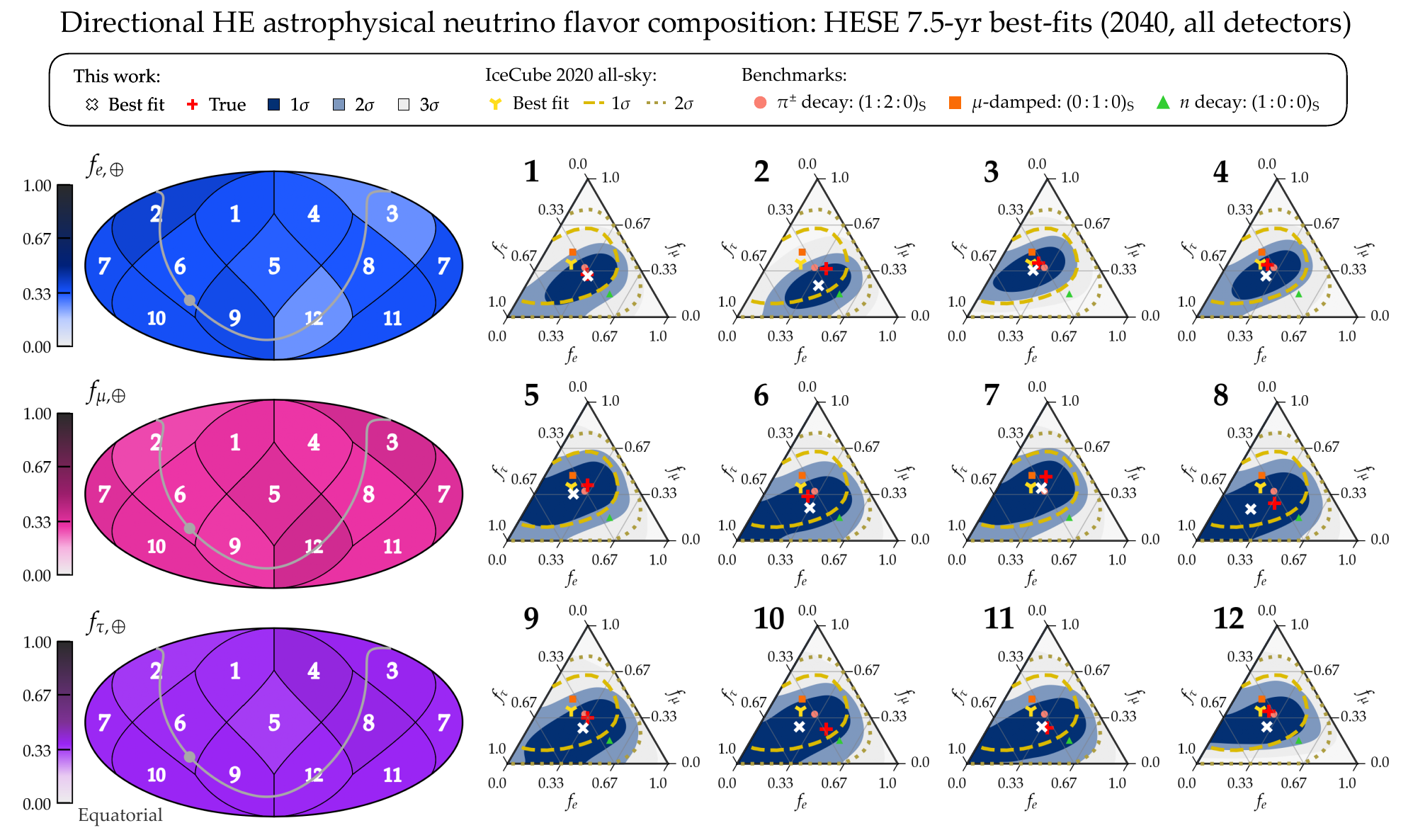}
  \vspace{-20pt}
  \caption{\textbf{\textit{Directional flavor composition, projections for 2040, assuming the best-fit flavor composition from present-day IceCube data.}}  Results are obtained using the cumulative, combined measurements of planned neutrino telescopes by 2040 (\figu{cl_limits}): Baikal-GVD, IceCube, IceCube-Gen2, KM3NeT, P-ONE, and TAMBO. \textit{The results are compatible with isotropy in all flavors.} }
  \label{fig:all_ternaries_aniso_2040_atbfs}
 \end{figure*}

Figure~\ref{fig:cl_disco} forecasts the potential to discover anisotropy in the flavor composition of high-energy astrophysical neutrinos.  It shows the mean minimum values of the dipole power, $C_{\ell = 1}^\alpha$, and the quadrupole power, $C_{\ell = 2}^\alpha$, that are distinguishable from isotropy, \ie, from $C_{\ell = 1}^\alpha = C_{\ell = 2}^\alpha = 0$, at a 90\%~C.L.  Our results are provided merely as indicative.  For $C_{\ell = 1}^\alpha$, the discovery potential is computed by allowing only the coefficients $q_{1,0}^e$, $q_{1,0}^\mu$, and $q_{1,0}^\tau$ to float, while, for $C_{\ell = 2}^\alpha$, they are computed by allowing only $q_{2,0}^e$, $q_{2,0}^\mu$, and $q_{2,0}^\tau$ to float. Figure~\ref{fig:cl_disco} complements Fig.~\ref{fig:cl_limits}, which shows the upper limits on $C_{\ell = 1}^\alpha$ and $C_{\ell = 2}^\alpha$. For the majority of our forecast timeline, the discoverable anisotropies lie outside the range already constrained by current data, but we show that potentially by 2035 for the dipole, and by 2040 for the quadrupole, flavor anisotropies that are currently allowed also become accessible to detection. 

Figure~\ref{fig:all_ternaries_aniso_2040_atbfs} complements \figu{all_ternaries_data} by simulating events driven by a flux with the current best-fit flavor harmonic moments, and reconstructing them with all detectors assumed to be in operation by 2040. As we simulate the harmonic moments, and not the exact pixel best-fit values, we observe that the ``true'' values are in some pixels slightly distanced from the HESE best fit values. This is an artifact of only simulating structures up to the size of a quadrupole, and therefore not capturing smaller angular structures. In this case, our results only demonstrate the future feasibility of discovering an anisotropy similar to the best fit of our current analysis. As the current uncertainties are large in every pixel, the exact values that we simulate do not change these conclusions. With the advent of all detectors in 2040, we observe a corresponding shrink in the flavor posteriors in each pixel. However, the overall directional flavor skymaps are still compatible with isotropy in all flavors. The evidence for anisoropy grows, but is not significant according to Jeffreys' criteria~\cite{jeffreys1998theory} ($\mathcal{B} = 2.3 \pm 0.4$).

\begin{figure*}[!htb]
  \centering
  \includegraphics[trim={0 0.3cm 0 0.25cm}, clip, width=1\columnwidth]{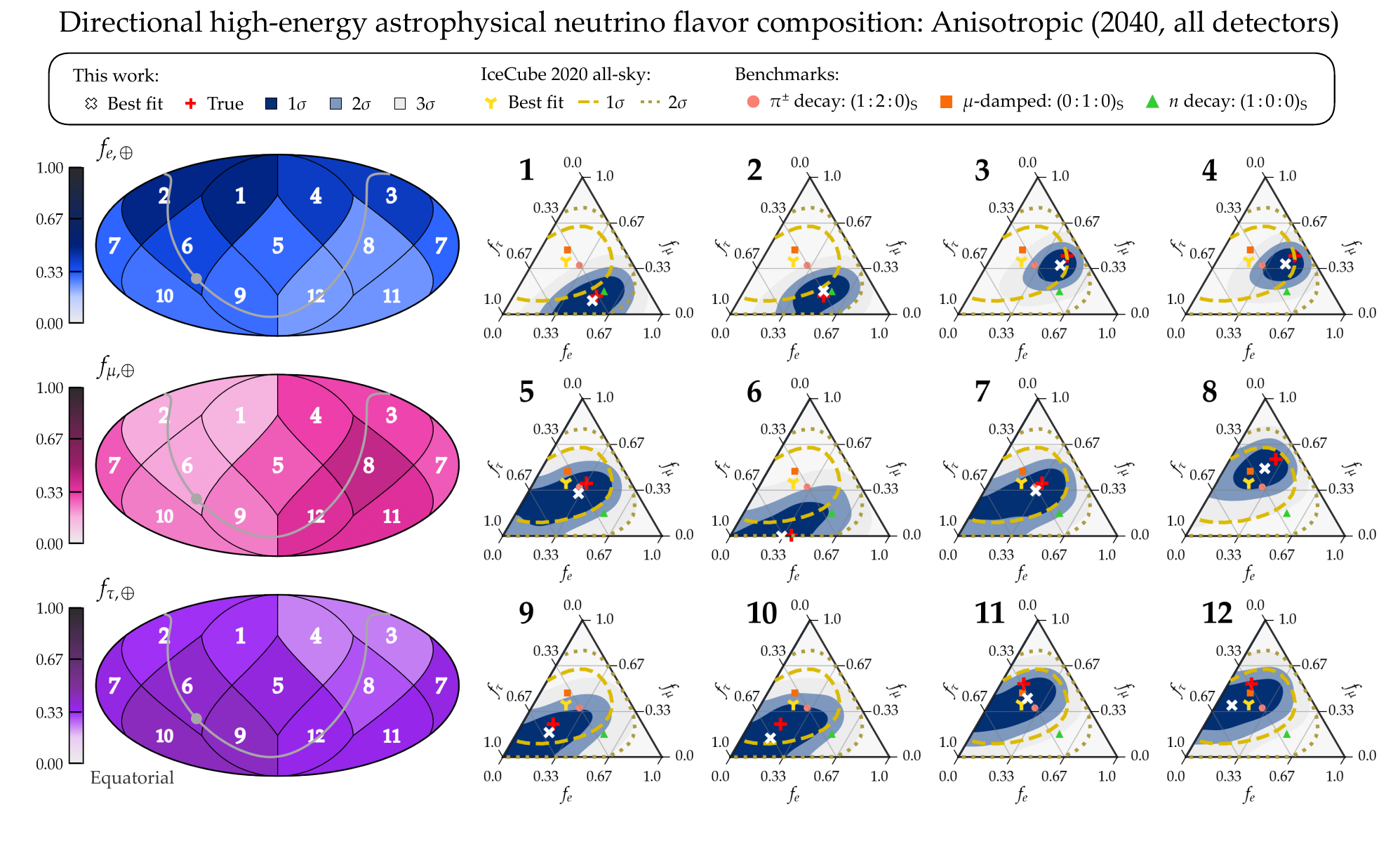}
  \vspace{-30pt}
  \caption{\textbf{\textit{Directional flavor composition, projections for 2040, assuming large flavor anisotropy.}} Results are obtained using the cumulative, combined measurements of planned neutrino telescopes by 2040 (\figu{future-detectors}): Baikal-GVD, IceCube (and IceCube Gen2), and KM3NeT for flavor fluxes simulated using the harmonic moments: $q^e_{1,0}=1.5, q^\mu_{1,0}=-0.5, q^\tau_{1,0} = -1$ and $q^e_{1,1} = 0.2i, q^\mu_{1,1} = -1.2i, q^\tau_{1,1} = -i$. \textit{Evidence for anisotropy is substantial.} }
  \label{fig:all_ternaries_aniso_2040_large}
 \end{figure*}

Figure~\ref{fig:all_ternaries_aniso_2040_large} shows the projections of the directional flavor composition for the year 2040, assuming large flavor anisotropy. The evidence for anisotropy is very strong according to Jeffreys' criteria ($\mathcal{B} = 32 \pm 4$). We can see that now, the reconstructed best-fit skymaps show clear flavor anisotropies, and the allowed regions in the ternary plots of each pixel show significant directional preference. 

\begin{figure*}[!htb]
  \centering
  \includegraphics[trim={0 0.3cm 0 0.25cm}, clip, width=1\columnwidth]{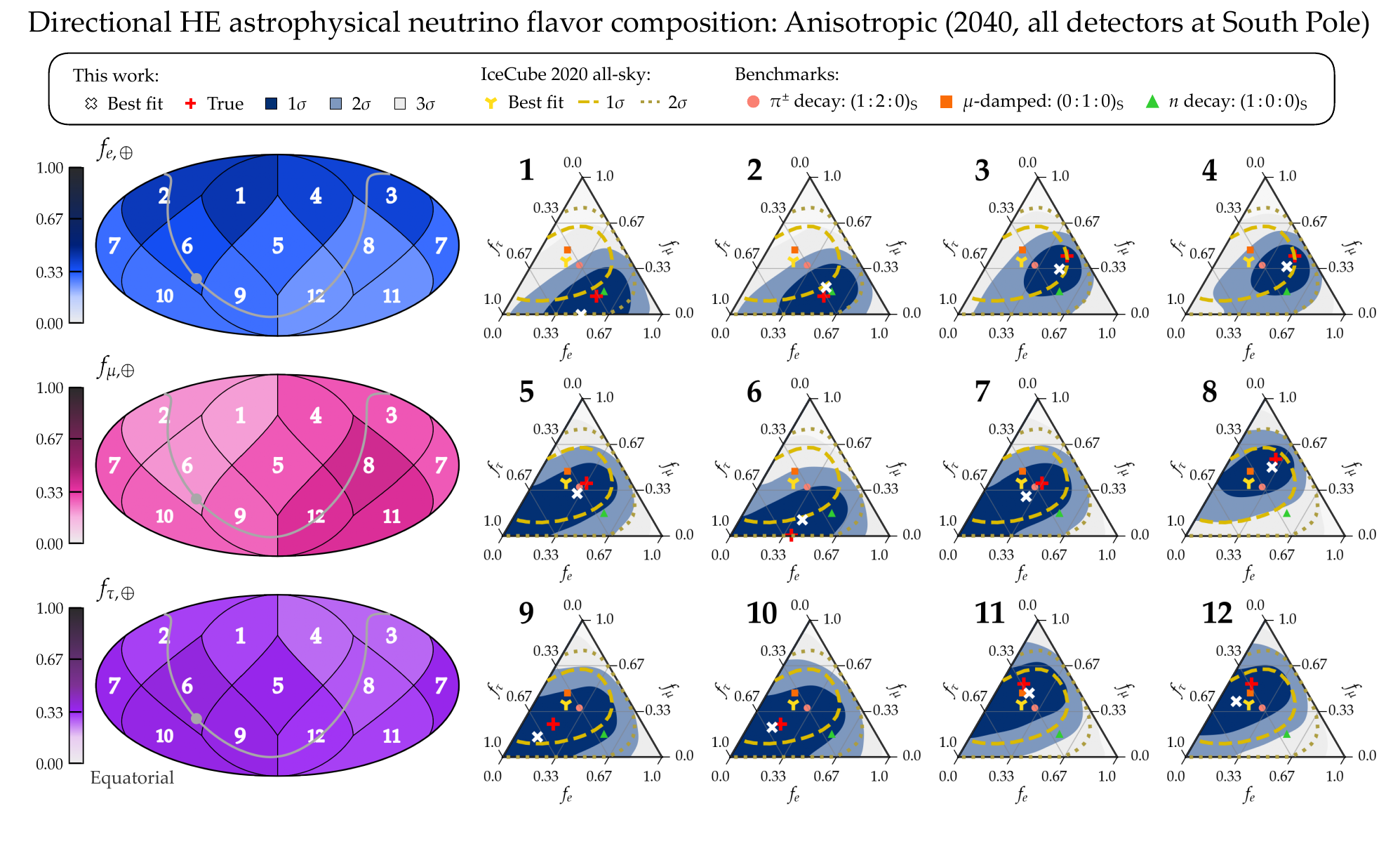}
  \vspace{-30pt}
  \caption{\textbf{\textit{Directional flavor composition, projections for 2040, assuming large flavor anisotropy and all detectors placed at the South Pole.}} Same as Fig.~\ref{fig:all_ternaries_aniso_2040_large}, but with detectors artificially placed at the South Pole. \textit{Evidence for anisotropy is strong.} }
  \label{fig:all_ternaries_aniso_2040_southpole}
 \end{figure*}

Figure~\ref{fig:all_ternaries_aniso_2040_southpole} simulates the same flux as \figu{all_ternaries_aniso_2040_large}, but the posteriors are recovered after we artificially placed all upcoming detectors at the South Pole, to show the gain from spreading them out.  Relative to \figu{all_ternaries_aniso_2040_large}, for which the detectors are placed at their actual planned locations (\figu{future-detectors}), the evidence for anisotropy degrades appreciably ($\mathcal{B} = 11 \pm 2$) compared to having globally-distributed detectors. 

In Appendix~\ref{sec:detailed_results}, we show the full posterior corner plots for all our fitting parameters, as well as the posteriors of the harmonic moments we recover from the flavor skymaps.

\subsection{Using flavor skymaps to constrain Lorentz-invariance violation}
Lorentz invariance, a pillar of the Standard Model, might be broken at high energies, including for neutrinos~\cite{Colladay:1998fq, Mattingly:2005re}.  If so, the presence of Lorentz-invariance violation (LIV) should become more evident in high-energy astrophysical~\cite{Arguelles:2015dca, Bustamante:2015waa, IceCube:2021tdn} (and atmospheric~\cite{IceCube:2017qyp}) neutrinos.  To describe LIV, we use the Standard Model Extension~\cite{Colladay:1998fq, Kostelecky:2003cr, Kostelecky:2008ts, Diaz:2011ia}, an effective field theory that couples neutrinos to spacetime features that define preferred directions.  Most searches for LIV in neutrinos look for isotropic or sidereal effects induced by the translation of the Earth.  Searches for persistent misalignments with a preferred direction---known as ``compass asymmetries''~\cite{Kostelecky:2003cr}---where neutrino oscillations depend on the neutrino trajectory, are comparatively poorly explored.  Our new sensitivity to high-energy neutrino flavor anisotropies allows us to mend this. Our approach is outlined in detail in Appendix~\ref{sec:liv}; here, we give a breif summary of the preliminary results.

We illustrate the utility of using the flavor skymaps to investigate new physics by constraining one of the many flavor anisotropy-generating LIV operators. Here, we treat  dimension-5, CPT-violating LIV couplings, whose effects grow $\propto E^2$~\cite{Kostelecky:2008ts, Kostelecky:2011gq}.  Current constraints on the parameters that drive compass asymmetries come from GeV accelerator neutrinos, $( \hat{a}_{\rm eff}^{(5)} )_{\ell,m}^{\alpha,\beta} \lesssim 10^{-20}$--$10^{-15}$~GeV$^{-1}$~\cite{LSND:2005oop, Kostelecky:2008ts, MiniBooNE:2011pix, Kostelecky:2011gq}, where $\ell$ and $m$ denote the expansion of the LIV operators in spherical harmonics.

{The present lack of evidence for flavor anisotropy in high-energy astrophysical neutrinos already improves limits by $\sim$$10^{15}$, \eg, $\lvert {\rm Re}(\hat{a}_{\rm eff}^{(5)})^{e,\mu}_{2,1} \rvert < 2.69 \cdot 10^{-33}$ GeV$^{-1}$ at $2\sigma$. By 2030, this could become $\lvert {\rm Re}(\hat{a}^{(5)}_{\rm eff})^{e,\mu}_{2,1} \rvert < 2.45 \cdot 10^{-33}$ GeV$^{-1}$; by 2040, $8.86 \cdot 10^{-34}$ GeV$^{-1}$.  We leave in-depth study to future work.}

\section{Discussion and Summary}

For the first time, we have measured the directional dependence of the flavor composition of high-energy astrophysical neutrinos, \ie, the distributions of arrival directions of $\nu_e$, $\nu_\mu$, and $\nu_\tau$.  This new observable combines proven capabilities of present and future high-energy neutrino telescopes to enable new tests of astrophysics and particle physics.  

Using present IceCube observations (7.5~years of HESE), we find that the skymaps of $\nu_e$, $\nu_\mu$, and $\nu_\tau$ are compatible with isotropy (\figu{all_ternaries_data}). {We quantify this as limits on the angular power in flavor dipoles and quadrupoles, and constrain the existence of flavor compass asymmetries induced by Lorentz-invariance violation.  Our forecasts (Figs.~\ref{fig:all_ternaries_aniso_2040_atbfs}--\ref{fig:all_ternaries_aniso_2040_southpole})}, built on the combined detection by multiple planned neutrino telescopes up to the year 2040, show clear, though cautious, improvement.

Our measurement lies at the limit of what is possible with present experimental capabilities.  While being conservative in our assumptions, we envision opportunities for improvement.  Today, the main limitations are the poor angular resolution of HESE cascades and the difficulty in identifying flavor.  In the future, these could be overcome by progress in event reconstruction~\cite{IceCube:2021umt, IceCube:2022njh, IceCube:2023ame}; including through-going muons~\cite{IceCube:2021uhz}, abundant and with better angular resolution but higher atmospheric contamination; and using new techniques in flavor identification, like dedicated templates~\cite{IceCube:2020fpi, IceCube:2023fgt} and muon and neutron echoes~\cite{Li:2016kra, Steuer:2017tca, Farrag:2023jut}.  

A decade into the discovery of high-energy astrophysical neutrinos, challenging measurements that were formerly out of reach start to become possible.  As we move forward, pursuing them will push the limits of our insight.

\section{Acknowledgements}
MB and BT are supported by {\sc Villum Fonden} under project no.~29388.  This work used resources provided by the High-Performance Computing Center at the University of Copenhagen.

\clearpage
\appendix

\section{Method Validation}\label{app:method_validation}
\renewcommand{\theequation}{A\arabic{equation}}
\renewcommand{\thefigure}{A\arabic{figure}}
\renewcommand{\thetable}{A\arabic{table}}
\setcounter{figure}{0} 
\setcounter{table}{0} 

The validity of our results rests on the ability of our methods to accurately infer anisotropy, or lack thereof, in the flavor composition of the diffuse flux of high-energy astrophysical neutrinos, based on the angular distribution of detected HESE events.  Below, we validate this ability.  Via mock event samples, we study how accurately the flavor-composition anisotropy can be inferred under different scenarios of isotropy and anisotropy in the flavor composition and in the all-flavor flux.  


\subsection{Present limits on all-flavor anisotropy}
\label{sec:valid_methods-present_limits_all_flavor}

In our directionally-dependent flavor flux model, we assumed that the all-flavor neutrino flux was isotropic at large angular scales. To test to which degree this is the case currently, we start by testing whether, based on the 7.5-year IceCube HESE sample, the measured all-flavor flux is compatible with its being isotropic. To do this, we fit the model in \equ{tot_aniso_evt_rate} to the HESE 7.5 year data.

\begin{figure}[h!]
 \centering
 \includegraphics[width=0.75\textwidth]{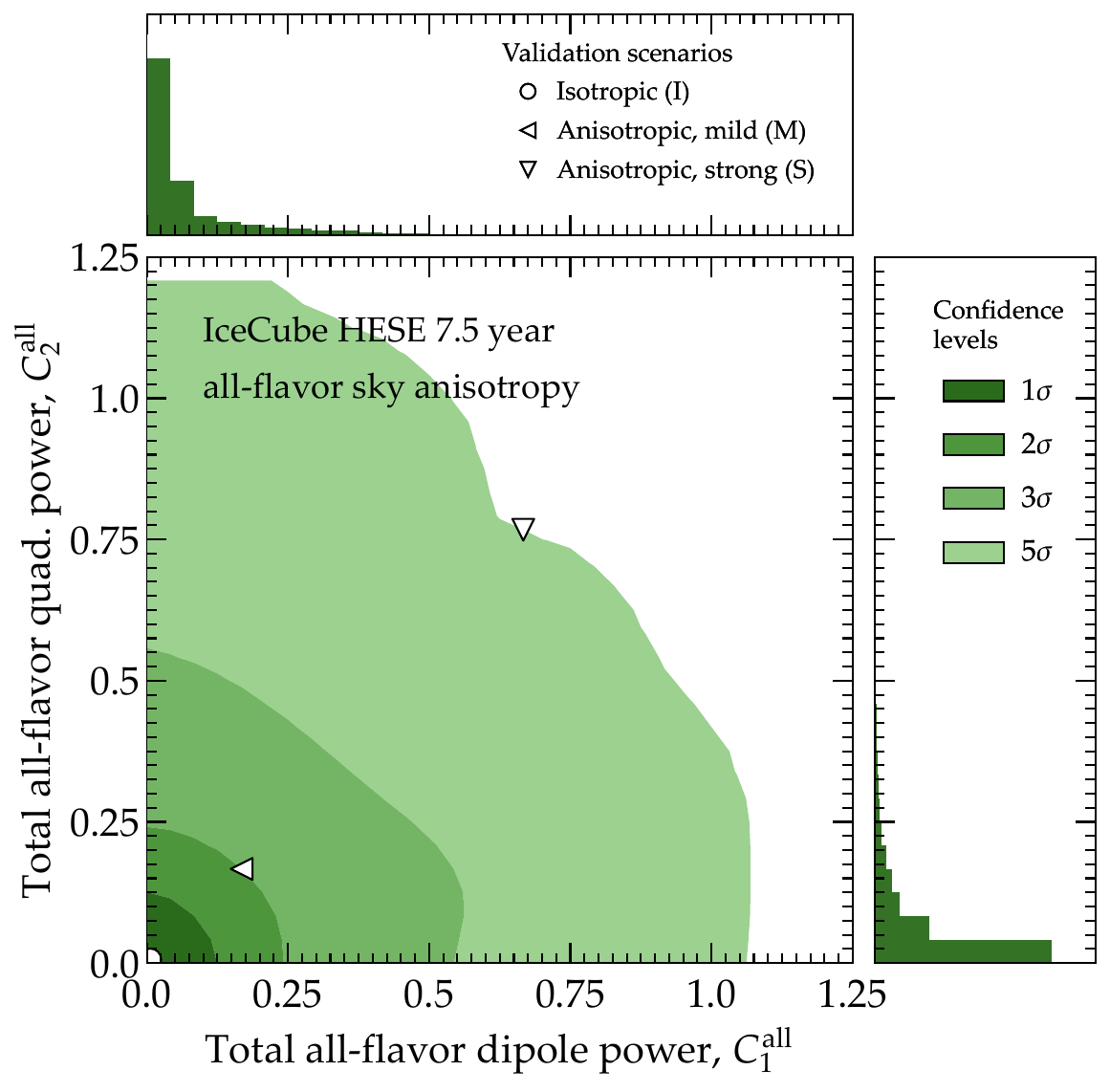}
 \caption{\textbf{\textit{All-flavor dipole and quadrupole power and their cross-correlation in the IceCube HESE 7.5-year data.}} Posteriors were recovered using the fitting model in \equ{tot_aniso_evt_rate} directly to the event rate skymaps, and isotropy was assumed for all flavor composition. The contours show the $1,2,3$ and $5\sigma$ confidence levels.} 
 \label{fig:tot_aniso_dat}
\end{figure}

Figure \ref{fig:tot_aniso_dat} shows the joint posterior distribution of the resulting harmonic powers, $C_{\ell = 1}^{\rm all}$ and $C_{\ell = 2}^{\rm all}$, that would capture dipole and quadrupole anisotropies of the all-flavor flux.  Their best-fit values are zero, corresponding to a best-fit isotropic all-flavor flux. To within $1\sigma$ or $2\sigma$, our results also indicate small allowed anisotropies away from the isotropic best-fit. Our results show that there is no large-scale all-flavor anisotropy in the present-day HESE data, in agreement to the results reported by the IceCube Collaboration~\cite{IceCube:2020wum}.  However, this does not necessarily preclude the presence of small-scale all-flavor anisotropy---which cannot be probed due to the large angular uncertainty of HESE events---nor of anisotropy in the flavor composition---which is the subject of our work.  Below, we build our validation scenarios based on \figu{tot_aniso_dat}.

However, any presence all-flavor anisotropy at large angular scales could bias the fits of the flux model we used. We therefore tested the performance of our model in \equ{tot_aniso_evt_rate} in the presence of both flavor-dependent and all-flavor anisotropies. 

\subsection{Validation scenarios and results}

In each validation scenario, we generate the mock observed event samples, $N_{{\rm obs}, i}^t$, using a flux prescription that allows for anisotropy in both the flavor composition and in the all-flavor flux (``fash'' for ``flavor and all-flavor spherical-harmonics''), \ie,
\begin{eqnarray}
 \label{equ:flux_and_flav_aniso}
 \Phi_{\nu_\alpha}^{\rm fash}
 (E, \theta_z, \phi, \boldsymbol{\theta}^{\rm fash})
 =
 \left( \frac{E}{100~{\rm TeV}} \right)^{-\gamma}
 && \times~
 \frac{1}{6}
 \left[
 1
 +
 \sum_{\ell=1}^\infty
 \sum_{m = -\ell}^\ell 
 q^\alpha_{\ell, m}
 Y^m_\ell(\theta_z, \phi)
 \right] 
 \nonumber \\
 && \times~
 \Phi_0
 \left[
 1
 +
 \sum_{\ell^\prime=1}^\infty 
 \sum_{m^\prime = -\ell^\prime}^{\ell^\prime} 
 b_{{\ell^\prime}, {m^\prime}}
 Y^{m^\prime}_{\ell^\prime}(\theta_z, \phi)
 \right] 
 \;,
\end{eqnarray}
where $\boldsymbol{\theta}^{\rm fash} \equiv (\Phi_0, \gamma, \{ q^\alpha_{\ell, m} \}, \{ b_{\ell, m} \})$ are the flux parameters, and $q^\alpha_{\ell, m}$ and $b_{\ell, m}$ are, respectively, the expansion coefficients of the flavor composition and of the all-flavor flux, which we set to different values depending on the validation scenario (Table~\ref{tab:validation}).  We set $\Phi_0$ and $\gamma$ to their best-fit values from the IceCube HESE analysis~\cite{IceCube:2020wum}.  The flavor-composition coefficients satisfy $\sum_{\alpha} q^\alpha_{\ell,m} = 0\ \forall\ \ell, m$, from demanding that $f_{e, \oplus} + f_{\mu, \oplus} + f_{\tau, \oplus} = 1$.  Like in Section~\ref{sec:stats_simulations}, we set the background parameters, $\boldsymbol{\eta}$, to their present-day best-fit values~\cite{IceCube:2020wum}.

Table~\ref{tab:validation} shows the six validation scenarios that we explore, corresponding to three illustrative choices of the pair of $C_{\ell = 1}^{\rm all}$ and $C_{\ell = 2}^{\rm all}$ in \figu{tot_aniso_dat}.  These choices correspond to different possibilities of the all-flavor flux: an isotropic flux (scenarios I), a mildly anisotropic flux (scenarios M), where $C_{\ell = 1}^{\rm all}$ and $C_{\ell = 2}^{\rm all}$ are $2\sigma$ away from their null best-fit values, and a strongly anisotropic flux (scenarios S), where they are $5\sigma$ away.  For each, we explore separately the cases of flavor-composition isotropy (I-i, M-i, S-i) and anisotropy (I-a, M-a, S-a).  Table~\ref{tab:validation} contains the specific values of the spherical-harmonic coefficients used in each scenario, according to the flux prescription that we introduce below.  These validation tests are not intended to be exhaustive, but to provide an indication of the validity and limitations of our methods. 

For each choice of the expansion coefficients in the different validation scenarios (Table~\ref{tab:validation}), we generate the mean expected mock observed event sample and extract the posteriors of the recovered flavor expansion coefficients, $(q_{\ell, m}^\alpha)_{\rm rec}$.  The goal of our validation tests is to determine whether the coefficients thus extracted reflect solely the anisotropy of the flavor-composition skymaps or if they are biased by the presence of anisotropy in the all-flavor skymap.

\begin{figure*}[h!]
  \includegraphics[width=1\textwidth]{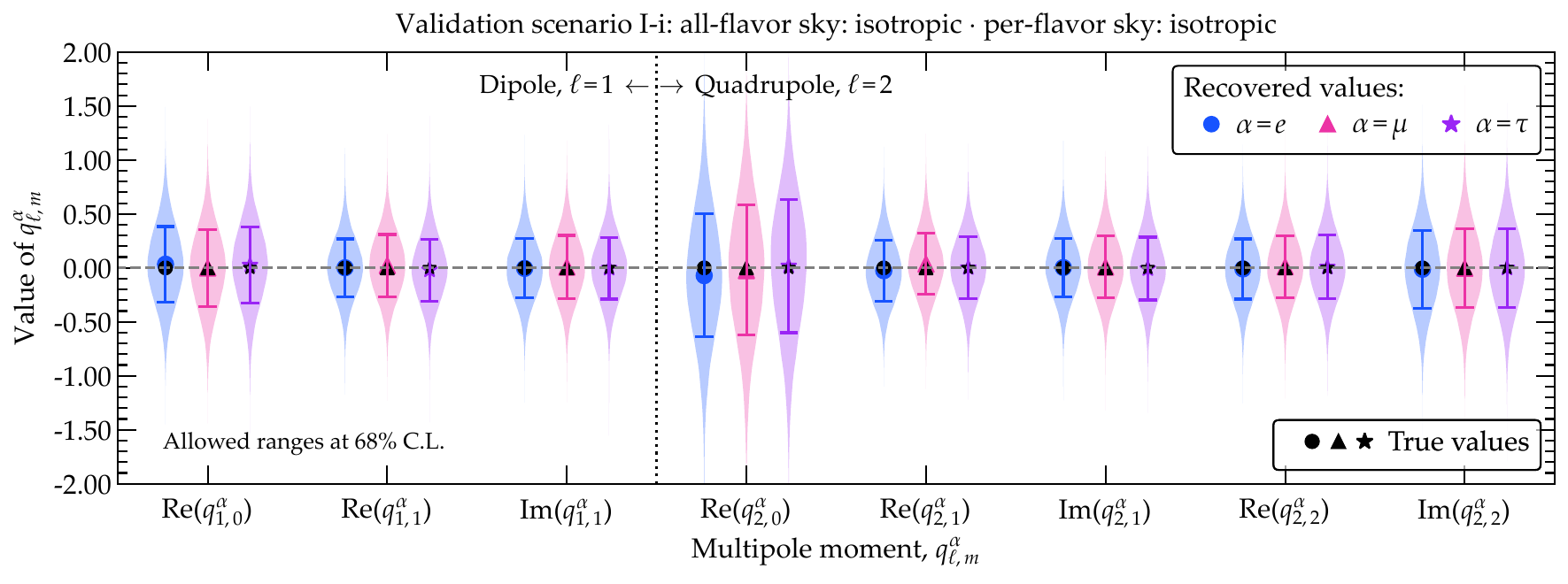}
  
 \includegraphics[width=1\textwidth]{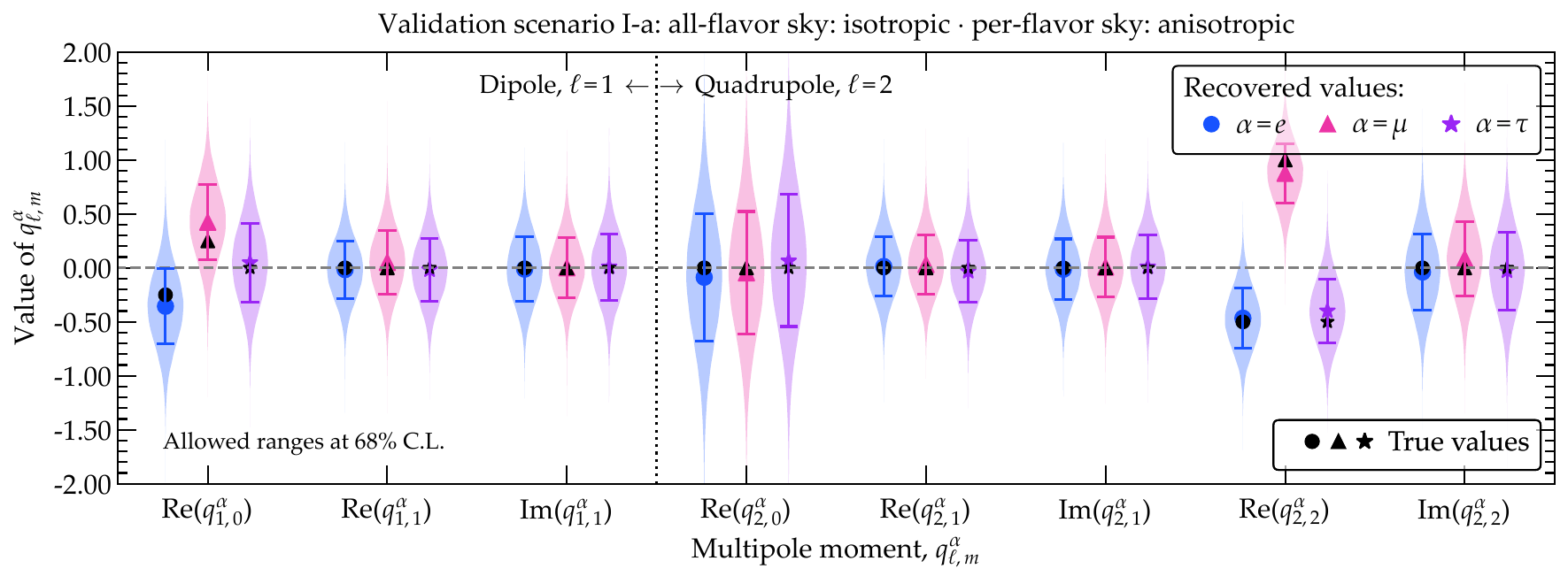}\vspace{-20pt}
 \caption{\label{fig:valid_sc_I}\textbf{\textit{Validation scenario I (isotropic all-flavor skymap): posteriors of the recovered flavor-anisotropy coefficients.}}  Recovery of the coefficients for flavor dipoles, $q^\alpha_{\ell=1,m}$ (\textit{left}), and quadrupoles, $q^\alpha_{\ell=2,m}$ (\textit{right}), of the spherical-harmonic expansion of the fluxes of neutrinos of different flavors, \equ{flav_flux_earth} in the main text.  \textit{Top:} Assuming the skymap of each flavor is isotropic (scenario I-i).  \textit{Bottom:} Assuming the skymap of each flavor is anisotropic, with the anisotropy given by Table~\ref{tab:validation} (scenario I-a).  See Table~\ref{tab:validation} for details.}
\end{figure*}

\begin{figure*}[t!]
  \includegraphics[width=1\textwidth]{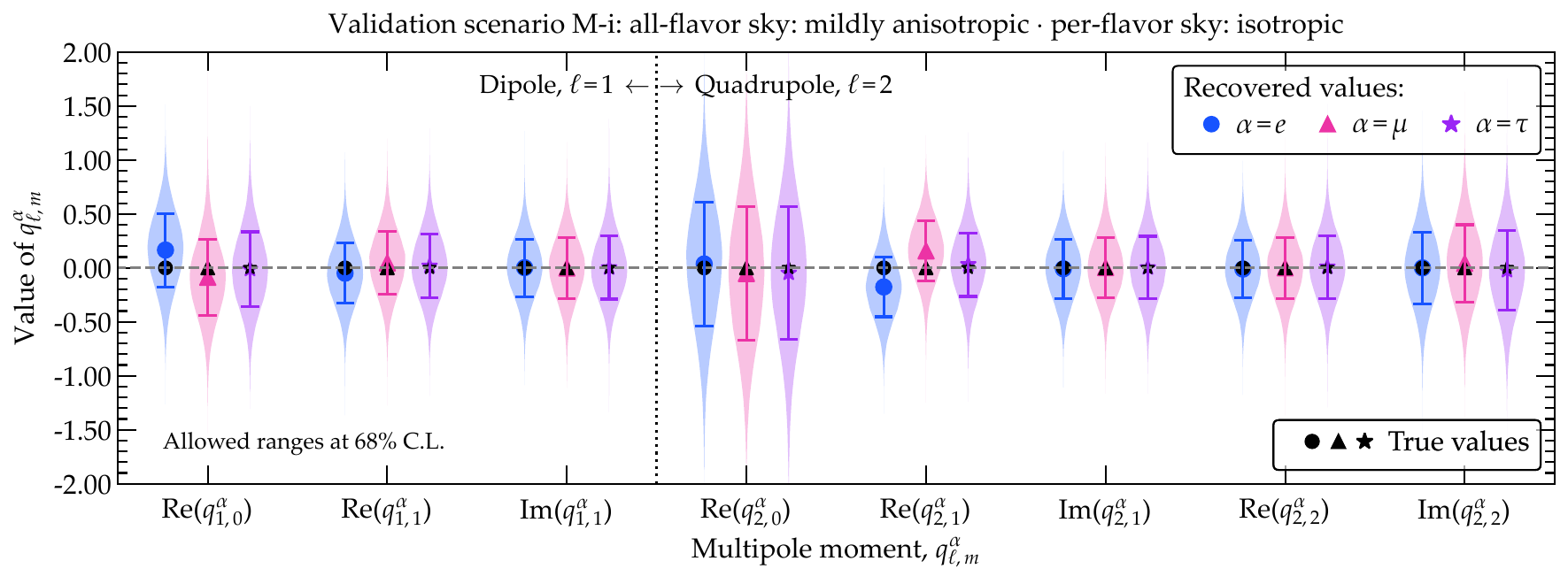}
  
 \includegraphics[width=1\textwidth]{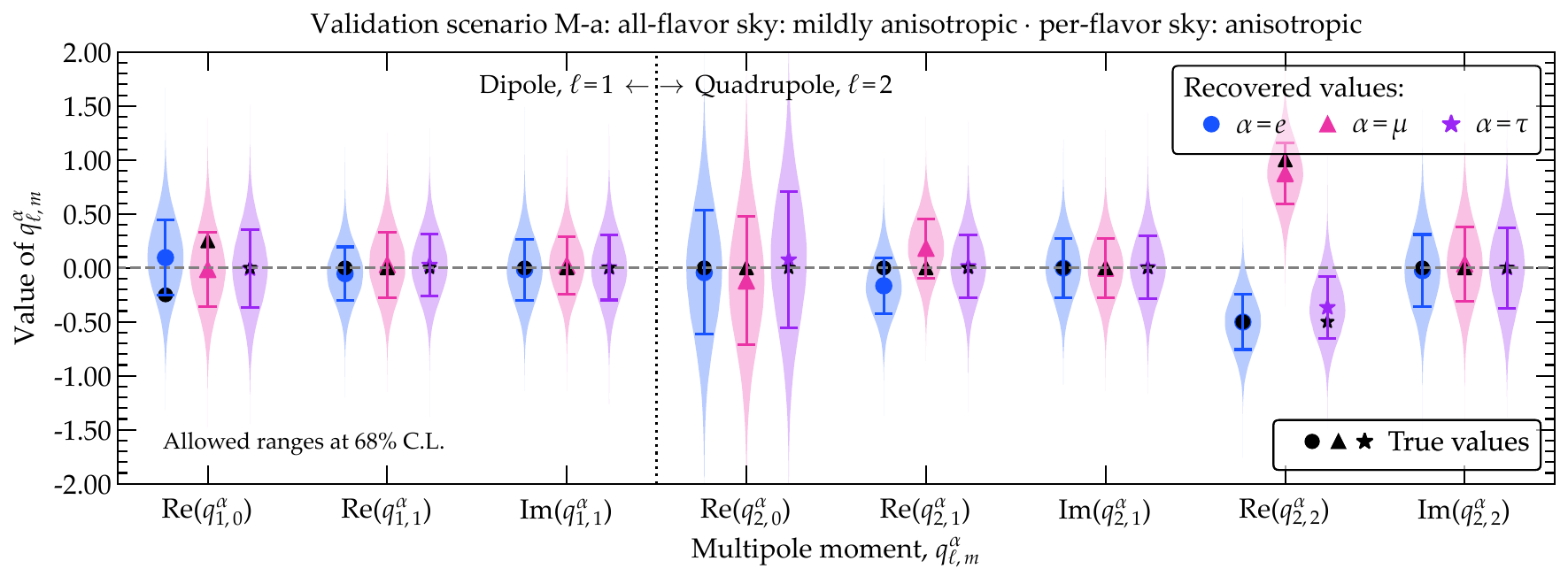}\vspace{-20pt}
 \caption{\label{fig:valid_sc_M}\textbf{\textit{Validation scenario M (mildly anisotropic all-flavor skymap): posteriors of the recovered flavor-anisotropy coefficients.}}  Same as \figu{valid_sc_I}, but for validation scenario M.  See Table~\ref{tab:validation} for details.}
\end{figure*}

\begin{figure*}[b!]
 \includegraphics[width=1\textwidth]{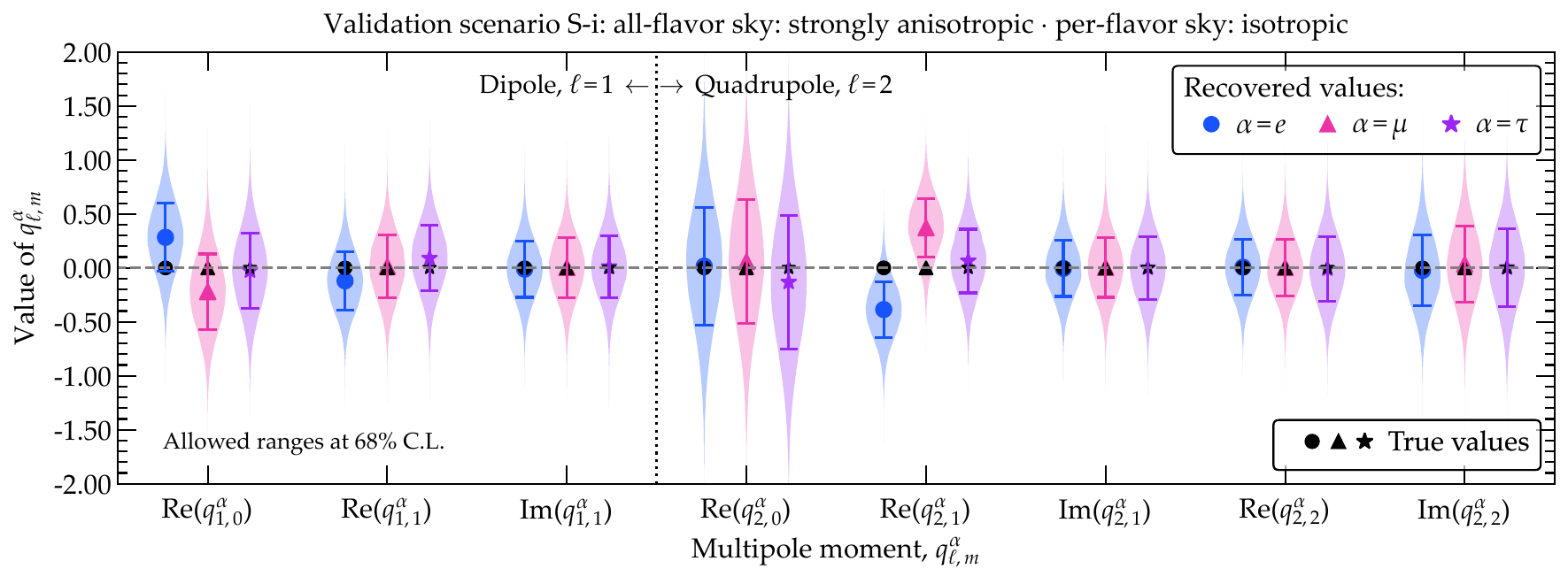}
 
 \includegraphics[width=1\textwidth]{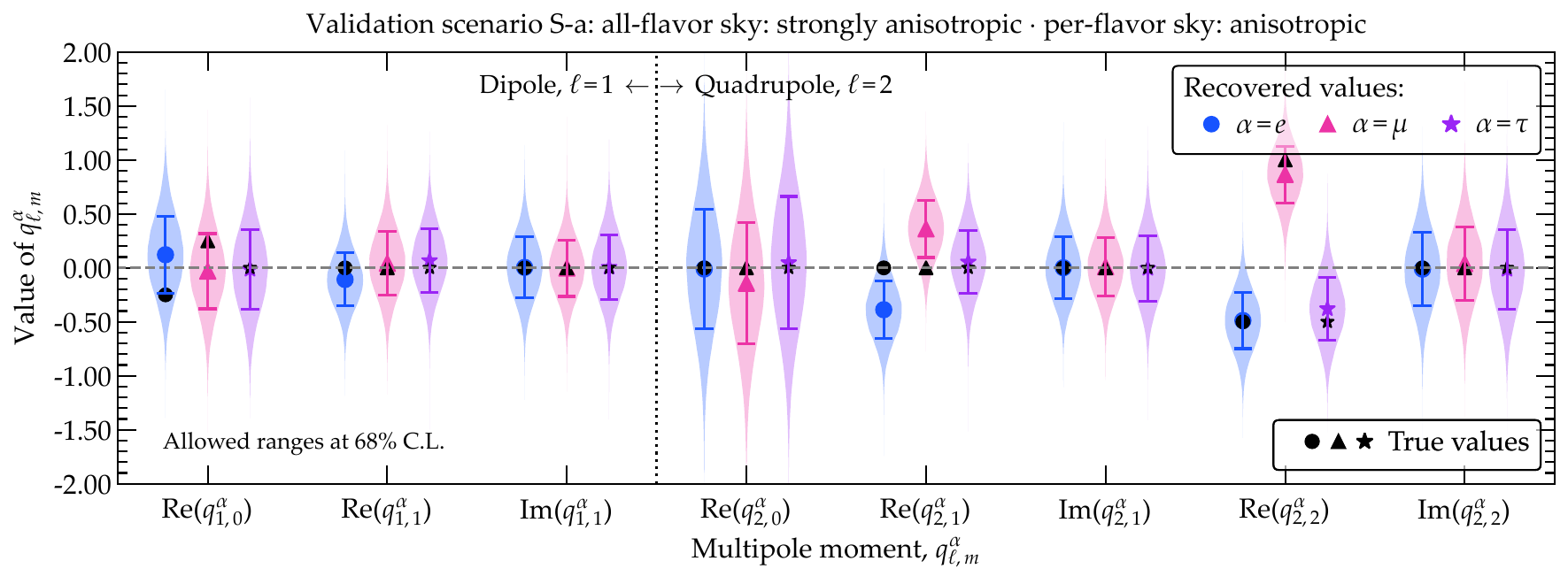}\vspace{-20pt}
 \caption{\label{fig:valid_sc_S}\textbf{\textit{Validation scenario S (strongly anisotropic all-flavor skymap): posteriors of the recovered flavor-anisotropy coefficients.}}  Same as \figu{valid_sc_I}, but for validation scenario S.  See Table~\ref{tab:validation} for details.}
\end{figure*}

Figures~\ref{fig:valid_sc_I}--\ref{fig:valid_sc_S} show the resulting posteriors of the flavor coefficients for all validation scenarios.  Table~\ref{tab:validation} summarizes the outcome.  We find that our methods accurately infer the isotropy or anisotropy in the flavor composition, regardless of whether the all-flavor flux is isotropic or not.  Scenarios M-i, M-a, S-i, and S-a show that large anisotropy (up to $5\sigma$ away from the present-day best fit) in the all-flavor flux may erode the accuracy in the measurement of the flavor composition.  However, even in these cases, the values of recovered flavor coefficients lie only within $\gtrsim 1\sigma$ of their true values.

\begin{table*}[h!]
  \caption{\label{tab:validation}\textbf{\textit{Validation scenarios of our procedure.}}  The validation scenarios test the recovery of the all-flavor and per-flavor coefficients of the spherical-harmonic expansion of the high-energy neutrino flux, 
  \equ{flux_and_flav_aniso}.  Validation scenarios are tested using event-rate projections for the year 2035, and assume detection only by IceCube.  Anisotropy in the all-flavor flux exists only in these validation tests; to produce our main results, we only allow for anisotropy in the flavor composition, while the all-flavor flux remains isotropic.\\}
  \centering
  \begin{tabular}{ccccc}
  \hline\hline
   \multirow{2}{*}{\makecell{Validation\\scenario}} & 
   \multirow{2}{*}{\makecell{All-flavor\\skymap}} & 
   \multirow{2}{*}{\makecell{Per-flavor\\skymap}} & 
   \multirow{2}{*}{Outcome of validation} &
   \multirow{2}{*}{Ref.} \\ \\
   \hline
   \multirow{1}{*}{I-i} &
   \multirow{1}{*}{Isotropic\textsuperscript{a}} &
   \multirow{1}{*}{Isotropic\textsuperscript{b}} &
   \multirow{2}{*}{\makecell{All-flavor and per-flavor\\coefficients recovered accurately}} &
   \multirow{2}{*}{\figu{valid_sc_I}} \\
   \multirow{1}{*}{I-a} &
   \multirow{1}{*}{Isotropic\textsuperscript{a}} &
   \multirow{1}{*}{Anisotropic\textsuperscript{c}} & 
   &
   \\ 
   \multirow{2}{*}{\makecell{M-i\\}} &
   \multirow{2}{*}{\makecell{Anisotropic,\\mild\textsuperscript{d}\\}} &
   \multirow{2}{*}{\makecell{Isotropic\textsuperscript{b}\\}} &
   \multirow{4}{*}{\makecell{Some leakage of all-flavor anisotropy\\onto per-flavor coefficients\\(values within $1\sigma$ of true values)}} &
   \multirow{4}{*}{\figu{valid_sc_M}} \\\\
   \multirow{2}{*}{\makecell{M-a\\}} &
   \multirow{2}{*}{\makecell{Anisotropic,\\mild\textsuperscript{d}}} &
   \multirow{2}{*}{\makecell{Anisotropic\textsuperscript{c}\\}} &
   \multirow{2}{*}{\makecell{ \\ }}&
   \\\\
   \multirow{2}{*}{S-i} &
   \multirow{2}{*}{\makecell{Anisotropic,\\strong\textsuperscript{e}}} &
   \multirow{2}{*}{Isotropic\textsuperscript{b}} &
   \multirow{4}{*}{\makecell{Larger leakage of all-flavor anisotropy\\onto per-flavor coefficients\\(values $\gtrsim 1\sigma$ away from true values)}} &
   \multirow{4}{*}{\figu{valid_sc_S}} \\\\
   \multirow{2}{*}{S-a} &
   \multirow{2}{*}{\makecell{Anisotropic,\\strong\textsuperscript{e}}} &
   \multirow{2}{*}{Anisotropic\textsuperscript{c}} &
   &
   \\\\\hline\hline
  \end{tabular}
\end{table*}\vspace{-10pt}
\noindent
\footnotesize{\textsuperscript{a}All all-flavor coefficients $b_{\ell, m} = 0$.\\
  \textsuperscript{b}All flavor-specific coefficients $q^\alpha_{\ell, m} = 0 \ (\alpha= e,\mu,\tau)$.\\
  \textsuperscript{c}Nonzero flavor-specific coefficients: $q^e_{1,0} = -0.25$, $q^\mu_{1,0} = 0.25$, $q^\tau_{1,0} = 0$, ${\rm Re}(q^e_{2,2}) = -0.5$, ${\rm Re}(q^\mu_{2,2}) = 1.0$, ${\rm Re}(q^\tau_{2,2}) = -0.5$.\\
  \textsuperscript{d}Nonzero all-flavor coefficients: $b_{1,0} = 0.63$, ${\rm Re}(b_{2,1}) = 0.80$.}\\
  \textsuperscript{e}Nonzero all-flavor coefficients: $b_{1,0} = 1.25$, ${\rm Re}(b_{2,1}) = 1.75$.

\clearpage

\section{Detailed Results}\label{sec:detailed_results}
\renewcommand{\theequation}{B\arabic{equation}}
\renewcommand{\thefigure}{B\arabic{figure}}
\renewcommand{\thetable}{B\arabic{table}}
\setcounter{figure}{0}
\setcounter{table}{0}

\begin{figure*}[t!]
 \centering 
 \includegraphics[height=1\textwidth]{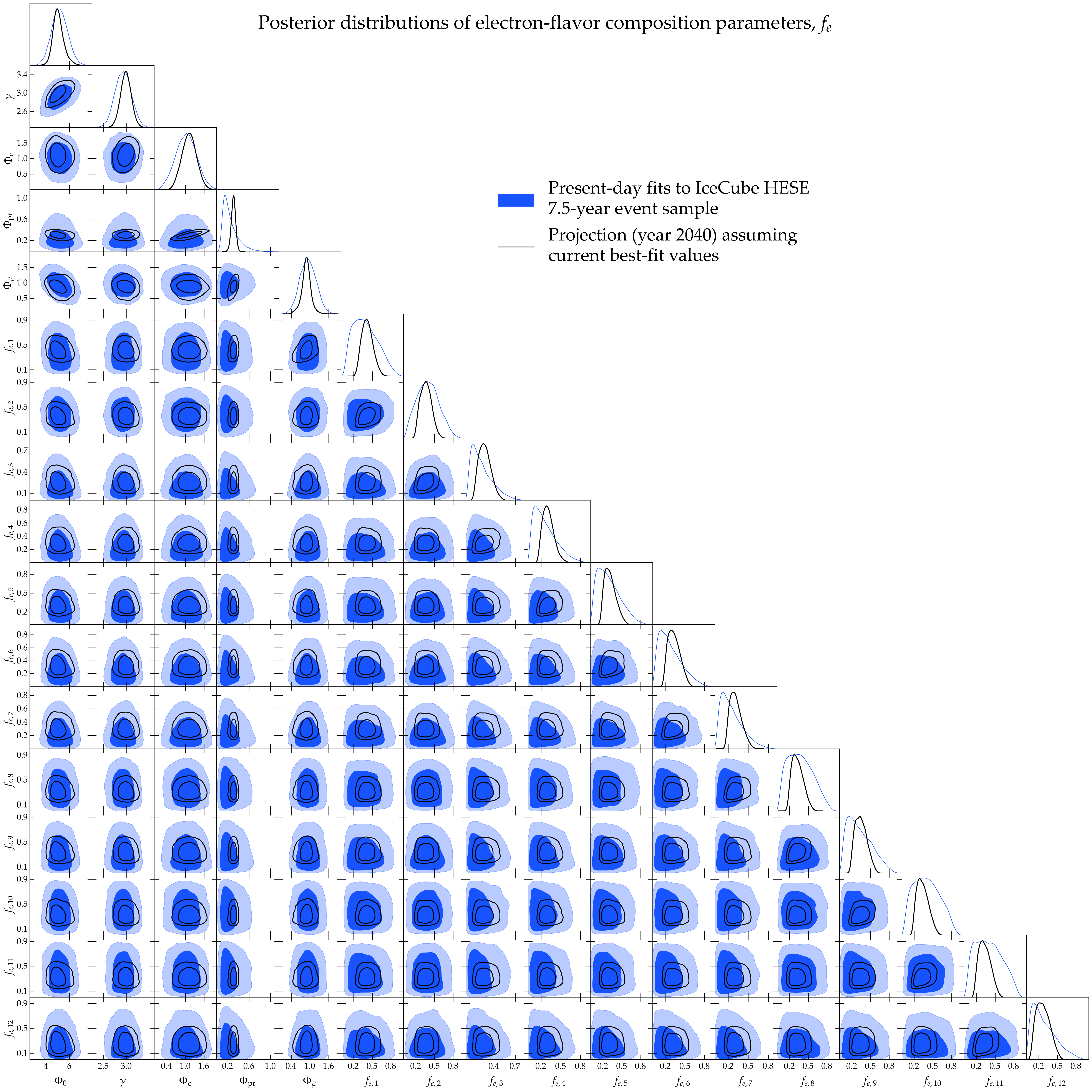}
 \caption{\textbf{\textit{Two-dimensional marginalized posterior distributions of the electron-flavor composition parameters, $\boldsymbol{f_{e, i }}$.}} Each panel shows the pairwise posterior distribution of two parameters, the result of marginalizing the full joint distribution over all the other parameters.  In each column, the top panel shows the one-dimensional posterior of each parameter.  See Table~\ref{tab:fit_params} for the corresponding one-dimensional allowed parameter ranges and \figu{cornerplot_mu} for the two-dimensional posterior distributions of the muon-flavor composition parameters, $f_{\mu, i}$.  See \figu{cornerplot_mu} for a similar plot for $f_{\mu, i}$.}
 \label{fig:cornerplot_e}
\end{figure*}
\begin{figure*}[t!]
 \centering
 \includegraphics[height=1\textwidth]{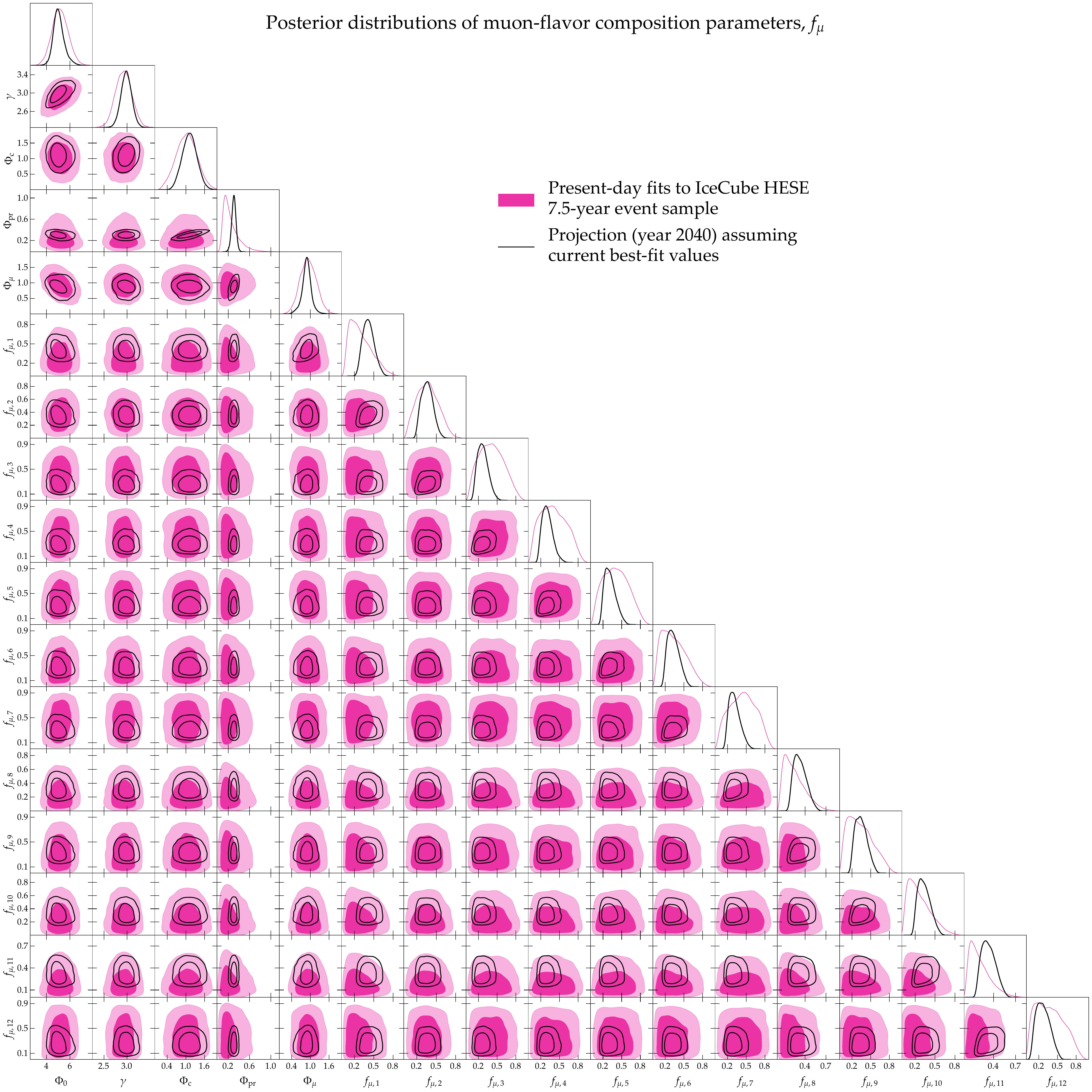}
 \caption{\textbf{\textit{Two-dimensional marginalized posterior distributions of the muon-flavor composition parameters, $\boldsymbol{f_{\mu, i}}$.}} Same as \figu{cornerplot_e}, but for $f_{\mu, i}$.}
 \label{fig:cornerplot_mu}
\end{figure*}

Figures~\ref{fig:cornerplot_e} and \ref{fig:cornerplot_mu} show the pairwise two-dimensional posterior distributions of the flavor-composition parameters, $f_{e,i}$ and $f_{\mu,i}$, respectively, and the other model parameters.  (The posterior distributions between $f_{e,i}$ and $f_{\mu,i}$ are shown instead as ternary plots in Fig.~\ref{fig:all_ternaries_data} and \ref{fig:all_ternaries_aniso_2040_atbfs})  There is little correlation between the flavor-composition parameters and the other parameters, which demonstrates that our results for $f_{e,i}$ and $f_{\mu,i}$ are not driven  by our choice of priors on $\Phi_0$, $\gamma$, $\Phi_\mu$, $\Phi_{\rm c}$, and $\Phi_{\rm pr}$.

\begin{figure*}[h!]
 \centering
 \includegraphics[width=\textwidth]{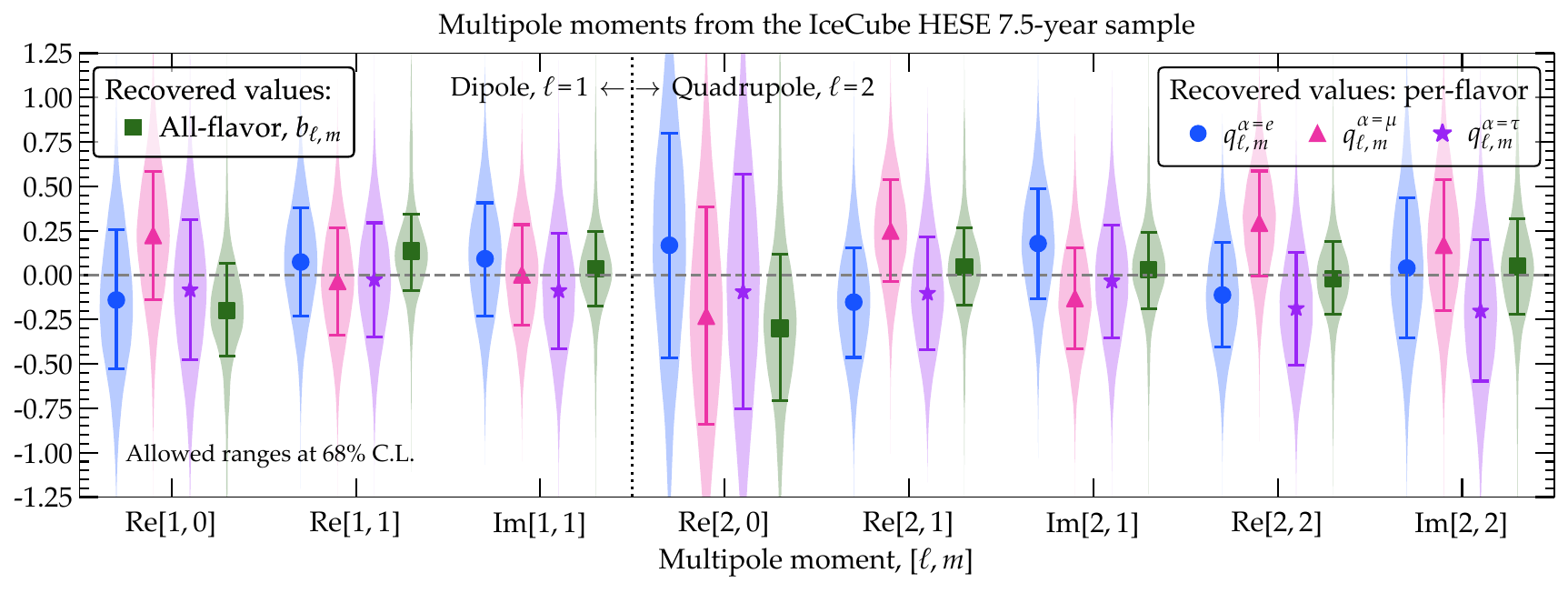}
 \caption{\textbf{\textit{One-dimensional marginalized posterior distributions of the spherical-harmonic coefficients of different flavors, $\boldsymbol{q^\alpha_{\ell,m}}$, and all-flavor, $\boldsymbol{b_{\ell,m}}$, inferred from a fit to the 7.5-year IceCube HESE sample.}}  The posteriors of $q^\alpha_{\ell,m}$ are inferred from the posteriors of the flavor-composition skymaps, $f_{e, i}$ and $f_{\mu, i}$, obtained from said fit (Figs.~\ref{fig:cornerplot_e}--\ref{fig:all_ternaries_data}).  Table~\ref{tab:harmonic_moment_data} contains numerical values of the best fits and allowed ranges.  Figure~\ref{fig:bfs_2040_multipoles} shows projected posteriors for the year 2040, assuming the present-day best-fit values as true. } 
 \label{fig:data_multipoles}
\end{figure*}

Figure~\ref{fig:data_multipoles} shows the one-dimensional posterior distributions of the flavor coefficients $q_{\ell,m}^\alpha$ ($\alpha = e, \mu, \tau$) of the dipole ($\ell = 1$) and quadrupole ($\ell = 2$) moments, inferred from the distributions of $f_{e, i}$ and $f_{\mu, i}$ obtained from the fit to the 7.5-year IceCube HESE sample.  In the fit, all of the coefficients are inferred simultaneously; \figu{data_multipoles} shows the one-dimensional posterior of each coefficient, computed by marginalizing the joint posterior probability distribution over all the other coefficients.  Figure~\ref{fig:data_multipoles} also shows results for the all-flavor coefficients, $b_{\ell, m}$, computed separately, under the assumption that the flavor skymaps are isotropic.  The best-fit values of all the coefficients are close to zero and are compatible with zero at the 68\%~C.L., signaling that there is no significant evidence of large anisotropy in present-day HESE data in the skymaps of the individual neutrino flavors and in the all-flavor skymap.  

\begin{figure*}[h!]
 \centering
 \includegraphics[width=\textwidth]{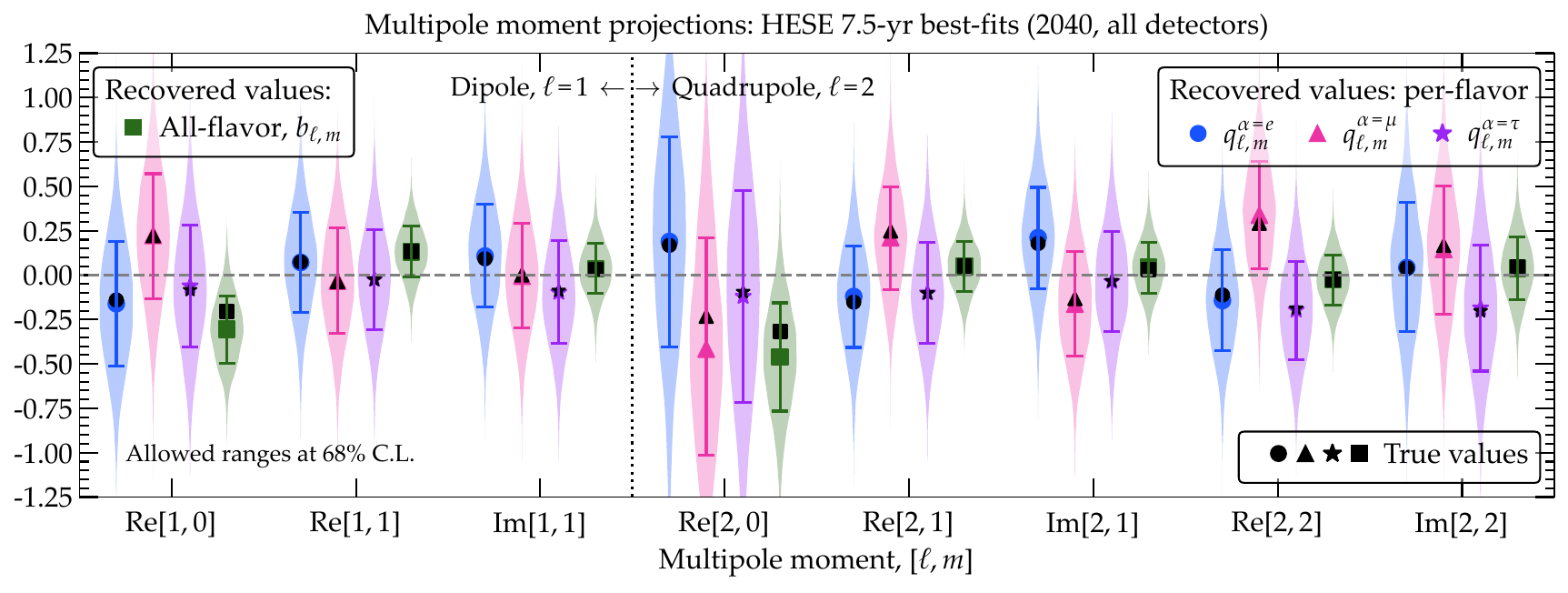}
 \caption{\textbf{\textit{One-dimensional marginalized posterior distributions of the spherical-harmonic coefficients of different flavors, $\boldsymbol{q^\alpha_{\ell,m}}$, and all-flavor, $\boldsymbol{b_{\ell,m}}$, inferred from projections for 2040, assuming the present best-fit flavor composition as true.}}  Same as Fig. \ref{fig:data_multipoles}, but true values are now the same as the best-fit values recovered from the HESE 7.5-year fits (Table~\ref{tab:harmonic_moment_data} and \figu{data_multipoles}). Results are obtained using the cumulative, combined measurements of planned neutrino telescopes by 2040 (\figu{cl_limits}): Baikal-GVD, IceCube, IceCube-Gen2, KM3NeT, P-ONE, and TAMBO.} 
 \label{fig:bfs_2040_multipoles}
\end{figure*}

Figure~\ref{fig:bfs_2040_multipoles} shows projections for the year 2040 of the posteriors of $q_{\ell,m}^\alpha$ and $b_{\ell, m}$, assuming their present-day best-fit values as their true values.  Compared to \figu{data_multipoles}, the posteriors are only marginally narrower, on account of the best-fit anisotropy being small (Table~\ref{tab:harmonic_moment_data}).

\clearpage
\section{Constraints on Lorentz-invariance violation}
\label{sec:liv}

\renewcommand{\theequation}{C\arabic{equation}}
\renewcommand{\thefigure}{C\arabic{figure}}
\renewcommand{\thetable}{C\arabic{table}}
\setcounter{figure}{0} 
\setcounter{table}{0} 
\subsection{Lorentz-invariance violation in neutrinos}
\label{sec:liv-formalism}

En route to Earth, high-energy astrophysical neutrinos oscillate in vacuum, driven by the standard Hamiltonian
\begin{equation}
 H_{\rm vac} 
 =
 \frac{1}{2E} 
 U_{\rm PMNS}M^2 U^\dagger_{\rm PMNS} \;,
\end{equation}
written in the flavor basis, where $E$ is the neutrino energy, $M^2 \equiv {\rm diag}(0,\Delta m_{21}^2, \Delta m_{31}^2)$ is the neutrino mass matrix, $\Delta m_{21}^2 \equiv m_2^2-m_1^2$, $\Delta m_{31}^2 \equiv m_3^2-m_1^2$, $m_i$ is the mass of the $\nu_i$ neutrino mass eigenstate, and $U_{\rm PMNS}$ is the Pontecorvo-Maki-Nakagawa-Sakata mixing matrix, parameterized by the mixing angles $\theta_{12}$, $\theta_{23}$, and $\theta_{13}$, and the CP-violation phase, $\delta_{\rm CP}$. 

In the Standard Model Extension (SME)~\cite{Kostelecky:2003fs, Kostelecky:2011gq}, the effective field theory that we adopt, Lorentz-invariance violation (LIV) in neutrinos is introduced via new couplings between neutrinos, which Lorentz-transform, and new fundamental tensors, which do not.  They introduce CPT-even and CPT-odd terms, and neutrino--antineutrino mixing, that affect neutrino oscillations in various ways.  In the main text and below, we focus on compass asymmetries~\cite{Kostelecky:2003cr}, \ie, the introduction via LIV of directional dependence into neutrino oscillations, otherwise nonexistent under standard oscillations in vacuum.  Because ours is the first exploration of compass asymmetries using high-energy astrophysical neutrinos, we illustrate our methods by focusing only on CPT-odd LIV.  This is driven by the Hamiltonian
\begin{equation}
 \label{equ:hamiltonian_liv}
 H_{\rm LIV}
 =
 \frac{1}{E}
 \hat{a}_{\rm eff} \;,
\end{equation}
where $\hat{a}_{\rm eff}$ is an effective CPT-odd operator~\cite{Kostelecky:2011gq} that is, in general, a function of the neutrino energy and momentum, and is represented as a $3 \times 3$ complex matrix with entries a priori unknown.  (The operator is called ``effective'' because it is a combination of other underlying operators.)  Below, we elaborate on it.

Thus, the total Hamiltonian that drives oscillations is
\begin{equation}
 \label{equ:hamiltonian_tot}
 H_{\rm tot} = H_{\rm vac} + H_{\rm LIV} \;.
\end{equation}
The relative sizes of $H_{\rm vac}$ and $H_{\rm LIV}$ vary with energy and depend on the values of the LIV coefficients (more on this later), and determine whether LIV effects are appreciable or not.  The flavor state $\nu_\alpha$ evolves in time, $t$, according to the Schr\"odinger equation, $i d\nu_\alpha /dt = H_{\rm tot} \nu_\alpha$, and this determines the $\nu_\alpha \to \nu_\beta$ flavor-transition probability ($\alpha, \beta = e, \mu, \tau$), $\nu_\beta^\dagger e^{-i H_{\rm tot} L} \nu_\alpha$, where the traveled distance $L \approx t$ because neutrinos are ultra-relativistic.  Later, in Appendix~\ref{sec:liv-constraints}, we compute our results using instead the average probability.

The LIV Hamiltonian is made up of contributions from operators of different dimension, $d$~\cite{Kostelecky:2011gq}.  We consider $d \geq 3$, which includes renormalizable and non-renormalizable contributions.  Further, following \Refe~\cite{Kostelecky:2011gq}, we expand the energy dependence of the LIV operator as a power series in neutrino energy---or, equivalently, in the absolute value of the neutrino momentum, $\boldsymbol{p}$---and its angular dependence as a spherical-harmonics series, \ie,
\begin{equation}
 \label{equ:aeff_series_exp}
 (\hat{a}_{\rm eff})^{\alpha,\beta} 
 =
 \sum_{d=3}^\infty
 \sum_{\ell=0}^{\ell_{\rm max}=d-1}
 \sum_{m=-\ell}^{\ell} 
 |\boldsymbol{p}|^{d-2} 
 Y_\ell^m(\hat{\boldsymbol{p}})(\hat{a}_{\rm eff}^{(d)})_{\ell,m}^{\alpha,\beta} \;,
\end{equation}
where $Y_\ell^m$ are complex spherical-harmonic functions and  $(\hat{a}_{\rm eff}^{(d)})_{\ell,m}^{\alpha,\beta}$ are the nine entries of the $3 \times 3$ matrix representation of the dimension-$d$ operator $(\hat{a}_{\rm eff}^{(d)})_{\ell,m}$.  They are scalar complex numbers, independent of the momentum, but dependent on the reference frame in which they are measured.  Following convention, we measure and report constraints on these LIV parameters in the reference frame centered on the Sun. 

\begin{figure}
 \centering
 \includegraphics[width=0.75\textwidth]{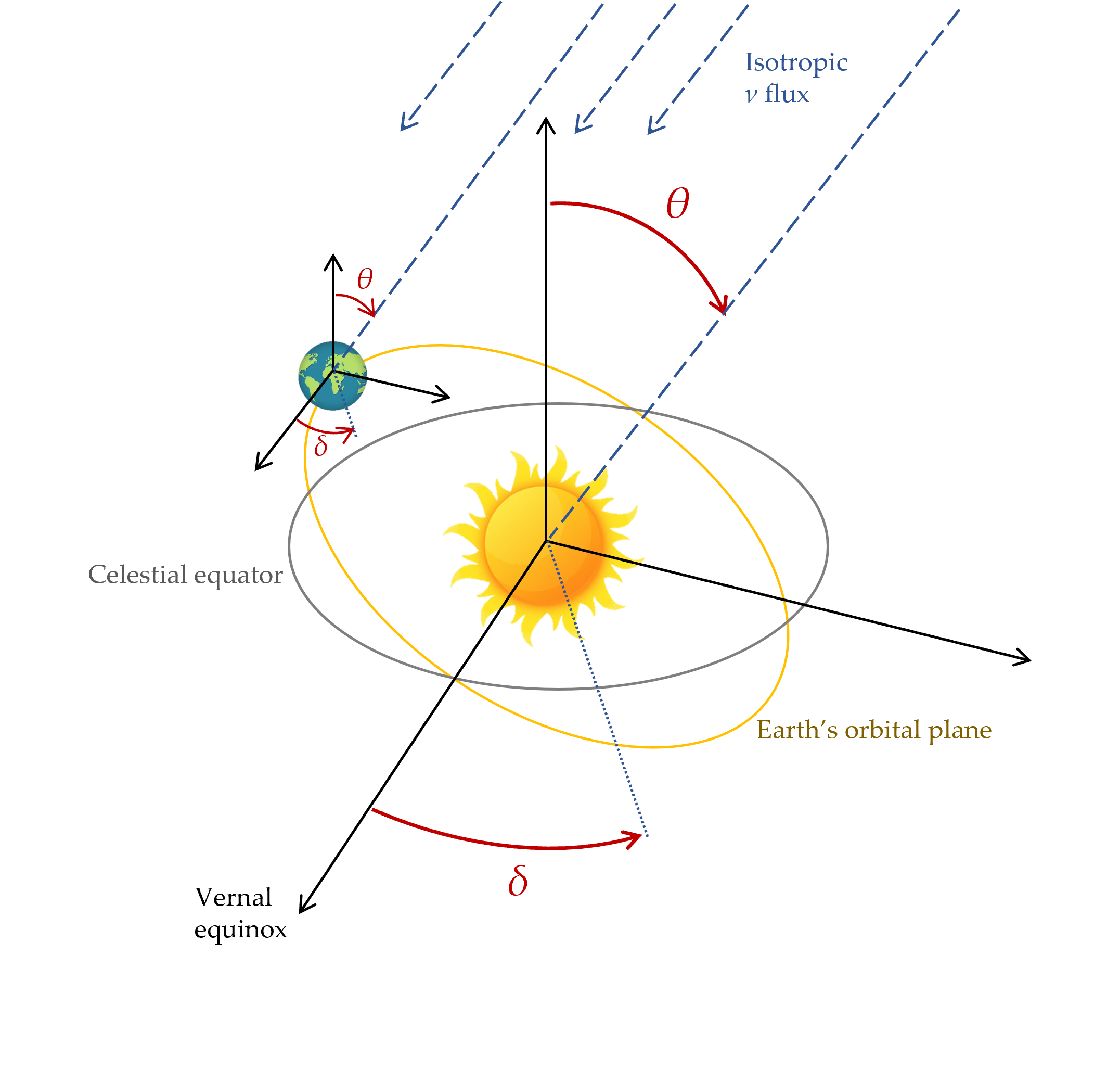}
 \caption{\textbf{\textit{Schematic diagram of the Sun-centered reference frame.}} {\color{black}We report constraints on the LIV coefficients expressed in the Sun-centered frame; they have the same values in the Earth-centered frame. The angles $\delta$ and $\theta$ define a particular direction in the sky, \ie, the arrival direction of a neutrino to Earth.  In our work, they are measured in a coordinate system centered on the Sun whose $x$-axis points from the center of the Sun to the vernal equinox (\ie, $\delta = 0^\circ$) and whose $z$-axis points  points along the rotation axis of the Earth (\ie, $\theta = 0^\circ$).  To a good approximation, the arrival direction is the same when measured in the Sun-centered frame and in the Earth-centered frame, since neutrinos originate at distances much greater than the distance between the Sun and the Earth. See Appendix \ref{sec:liv-formalism} for details.}}
 \label{fig:sun_frame}
\end{figure}

{\color{black}
Figure~\ref{fig:sun_frame} shows a schematic diagram of the Sun-centered reference frame, whose $z$-axis is aligned with the rotation axis of the Earth and whose $x$-axis points to the vernal equinox.  In the associated Earth-centered frame, where we measure the neutrino flavor composition, the neutrino arrival direction is parametrized in the equatorial coordinate system via an azimuth angle, $\delta$, where $\delta = 0^\circ$ defines the $x$-axis, and a polar angle, $\theta$, where $\theta = 0^\circ$ defines the $z$-axis (these, in practice, are the declination and right ascension coordinates, modulo conventions on the ranges of the angles).  Since we derive constraints on the LIV coefficients in the Earth-centered frame, we must transform them into the Sun-centered frame to report them.  Fortunately, in our case, the transformation between frames is trivial, as we explain below.

In general, the Lorentz boost of the neutrino four-momentum consists of an energy transformation and a spatial rotation.  Regarding the energy boost, the three-momenta in the Earth ($\oplus$) and Sun ($\odot$) frames are comparable, \ie, $|\boldsymbol{p}^{\oplus}| \approx |\boldsymbol{p}^\odot|$, since the orbital velocity of the Earth around the Sun is only about 0.01\% of the speed of light, so the neutrino energies are nearly the same in both frames.  Regarding the spatial rotation, there is none: because astrophysical neutrinos propagate over distances much greater than the separation between the Earth and the Sun, their arrival directions to the Earth and to the Sun are essentially the same, so there is no need to rotate between the frames.  Therefore, the limits on the LIV coefficients that we derive have the same values in the Earth-centered and Sun-centered frames.
}

\subsection{Constraints on LIV effective parameters}
\label{sec:liv-constraints}

The high-energy astrophysical neutrinos that are the subject of our work propagate, at least effectively, as an incoherent mixture of mass eigenstates, $\nu_1$, $\nu_2$, and $\nu_3$.  At TeV energies, the vacuum oscillation length is about $10^{-10}$~Gpc, negligible compared to the cosmological-scale baselines ranging from 100~Mpc to a few Gpc that these neutrinos are believed to travel.  Further, the energy resolution with which neutrinos are detected is typically $\gtrsim 10\%$.  As a consequence, neutrino telescopes are sensitive only to the average flavor-transition probability,
\begin{equation}\label{eq:average_osc_probability}
 P_{\nu_\alpha \to \nu_\beta}
 =
 \sum_{i=1}^3 |U_{\beta i}|^2|U_{\alpha i} |^2 \;,
\end{equation}
where $U$ is the matrix that diagonalizes the total Hamiltonian, $H_{\rm tot}$.  Given a flavor composition at the sources, $(f_{e, {\rm S}}, f_{\mu, {\rm S}}, f_{\tau, {\rm S}})$, the flavor composition at Earth is
\begin{equation}
 f_{\alpha, \oplus}
 =
 \sum_{\beta = e, \mu, \tau} 
 P_{\nu_\beta \to \nu_\alpha} f_{\beta, {\rm S}} \;.
\end{equation}

Via the vacuum oscillation Hamiltonian, $H_{\rm vac}$, the flavor composition depends on the values of the standard mixing angles, $\theta_{12}$, $\theta_{23}$, $\theta_{13}$, and the CP-violation phase, $\delta_{\rm CP}$.  Today, experimental uncertainties on their values are sizable and introduce uncertainty in the predicted flavor composition at Earth~\cite{Bustamante:2015waa, Song:2020nfh}.  In 2030 and 2040, the uncertainties in the mixing parameters are expected to shrink thanks to upcoming oscillation experiments DUNE, Hyper-Kamiokande, and JUNO, which would effectively make the uncertainty on the flavor composition at Earth vanish~\cite{Song:2020nfh}.  Hence, in our work, we have fixed the values of the standard mixing parameters to their present best-fit values from the recent NuFit 5.2 fit to global oscillation data, assuming normal neutrino mass ordering, without using Super-Kamiokande data~\cite{Esteban:2020cvm, NuFit5.2}.  Via the LIV Hamiltonian, $H_{\rm LIV}$, the flavor composition becomes also dependent on the values of the LIV parameters, $\boldsymbol{\lambda}$, on the neutrino energy and, for the compass asymmetries that we explore, on the direction of neutrino propagation, $\boldsymbol{\theta}$.  In summary, the flavor composition at Earth is $f_{\alpha, \oplus} \equiv  f_{\alpha, \oplus}(E, \boldsymbol{\theta}, \boldsymbol{\lambda})$.

The directional dependence introduced on the flavor-transition probability by LIV imprints itself on the otherwise isotropic diffuse neutrino fluxes of $\nu_e$, $\nu_\mu$, and $\nu_\tau$.  We compute the diffuse neutrino flux as the sum of contributions of identical high-energy neutrino sources, distributed in redshift, $z$.  We assume that the number density of sources, $\rho_{\rm src}$, is proportional to the star-formation rate, Eq.~(5) in \Refe~\cite{Yuksel:2008cu}, which peaks at $z \approx 1$.  This is representative of many different candidate classes of high-energy neutrino sources.  Each source injects a spectrum $\propto E^{-\gamma}$ of neutrinos.  We compute the flux of $\nu_\alpha$ as in Eq.~(B4) of \Refe~\cite{Bustamante:2016ciw}, \ie,
\begin{eqnarray}
 \label{equ:propagation}
 \Phi_\alpha(E, \boldsymbol{\theta}, \boldsymbol{\lambda})
 &=&
 \frac{\Phi_0}{E^2} \int_0^\infty
 dz~
 \frac{\rho_{\rm src}(z)}{h(z)(1+z)^2} \\
 && \times~
 [E(1+z)]^{2-\gamma} 
 f_{\alpha, \oplus}[E(1+z), \boldsymbol{\theta}, \boldsymbol{\lambda}]  
 \;, \nonumber 
\end{eqnarray}
where $\Phi_0$ is the normalization constant of the all-flavor flux, $h(z) = H_0\sqrt{\Omega_m(1+z)^3+\Omega_\Lambda}$ is the Hubble parameter, $H_0 = 100 h$~km~s$^{-1}$~Mpc$^{-1}$ is the Hubble constant, with $h = 0.674$, $\Omega_m$ = 0.315 is the adimensional energy density of matter, and $\Omega_\Lambda = 0.685$ is the adimensional energy density of vacuum~\cite{Planck:2018vyg}.  In practice, we integrate \equ{propagation} up to redshift of $z = 5$, since contributions from higher redshifts contribute negligibly.  The integrand in the left-hand side of \equ{propagation} is evaluated at an energy of $E(1+z)$ to account for the redshifting of the energy due to the adiabatic cosmological expansion.  When calculating constraints on LIV below, for illustration, we assume the flavor composition at the sources to be that from pion decay, $\left( \frac{1}{3}, \frac{2}{3}, 0 \right)_{\rm S}$, and the spectral index to be equal to the present IceCube HESE best-fit value, \ie, $\gamma=2.89$ \cite{IceCube:2020wum}.

\begin{figure*}[t!]
 \centering
 \includegraphics[width=1\textwidth]{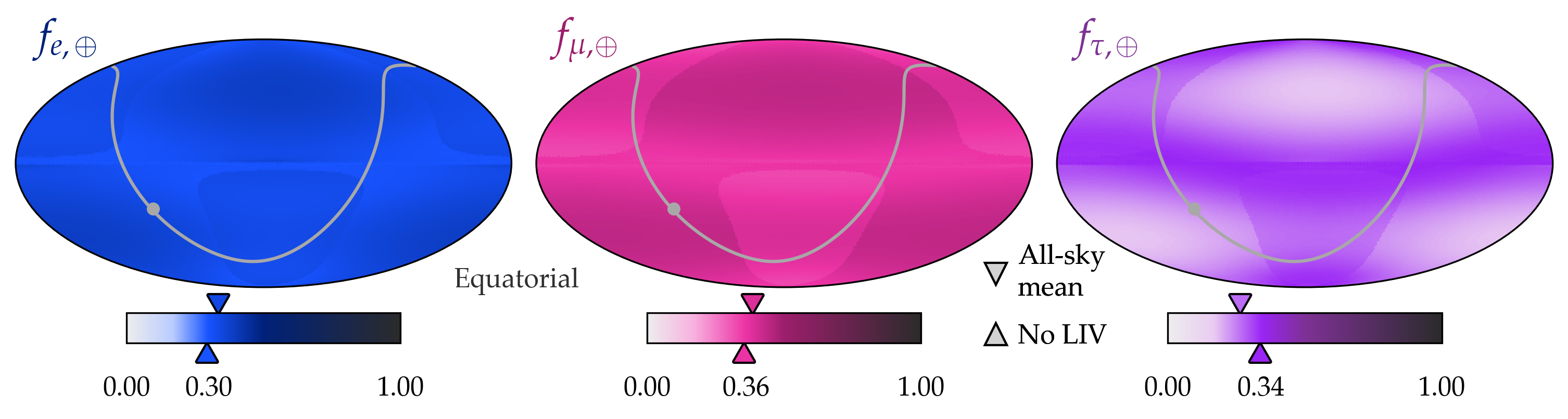}
 \caption{\textbf{\textit{{Example skymaps of flavor composition at Earth, $\boldsymbol{f_{\alpha, \oplus}}$, computed under LIV.}}}\textcolor{black}{The structures in the maps show flavor-dependent compass asymmetries introduced by LIV~\cite{Kostelecky:2003cr}.  These maps were generated using a single nonzero dimension-5 CPT-odd LIV parameter, its value saturating its present-day upper limit inferred from the IceCube HESE 7.5-year sample~\cite{IceCube:2020wum, IC75yrHESEPublicDataRelease}, ${\rm Re}(\hat{a}^{(5)}_{\rm eff})^{e,\mu}_{2,1} = 2.24 \times 10^{-33}$ GeV$^{-1}$, as described in Appendix~\ref{sec:liv-constraints}.  The flavor composition at the source is set to $\left( \frac{1}{3}, \frac{2}{3}, 0 \right)_{\rm S}$.  By construction, $f_{e, \oplus} + f_{\mu, \oplus} + f_{\tau, \oplus} = 1$ along every direction of the sky.  See Appendix \ref{sec:liv-formalism} for details.}}
 \label{fig:LIV_maps_example}
\end{figure*}

Figure~\ref{fig:LIV_maps_example} shows example skymaps of flavor composition at Earth under LIV, computed using \equ{propagation} using an illustrative value of ${\rm Re}(\hat{a}^{(5)}_{\rm eff})^{e,\mu}_{2,1}$.  They show that, indeed, the flavor composition is directionally modified and that the mean all-sky flavor composition is shifted from its value expected from standard oscillations.  The angular structures in \figu{LIV_maps_example} are a consequence of our particular choice of LIV, where only the $(\hat{a}_{\rm eff}^{(5)})_{2,1}^{e, \mu}$ coefficient of the CPT-odd LIV  operator $\hat{a}_{\rm eff}$ is nonzero.

The number of free LIV parameters is large, and their values are a priori unknown, to be determined by experiment.  For each operator dimension, $d$, the number of operators $(\hat{a}_{\rm eff}^{(d)})_{\ell,m}$ in the spherical-harmonics expansion of $\hat{a}_{\rm eff}^{(d)}$, \equ{aeff_series_exp}, is $\sum_{\ell = 0}^{\ell_{\rm max}} 2\ell+ 1 = d^2$, since $\ell_{\rm max} = d-1$.  Each of these operators is a $3 \times 3$ complex matrix, with each of its nine entries, $(\hat{a}_{\rm eff}^{(d)})^{\alpha,\beta}_{\ell,m}$, made up of real and imaginary parts.  However, because $\hat{a}_\text{eff}^{(d)}$ is Hermitian~\cite{Kostelecky:2011gq}, the entries satisfy $\left[ (\hat{a}_{\rm eff}^{(d)})^{\alpha,\beta}_{\ell,m} \right]^* = (-1)^m
(\hat{a}_{\rm eff}^{(d)})^{\beta,\alpha}_{\ell,-m}$, which halves the number of free parameters.  This totals $9d^2$ free parameters to determine for each operator dimension.  

\begin{figure}[h!]
 \centering
 \includegraphics[width=0.75\textwidth]{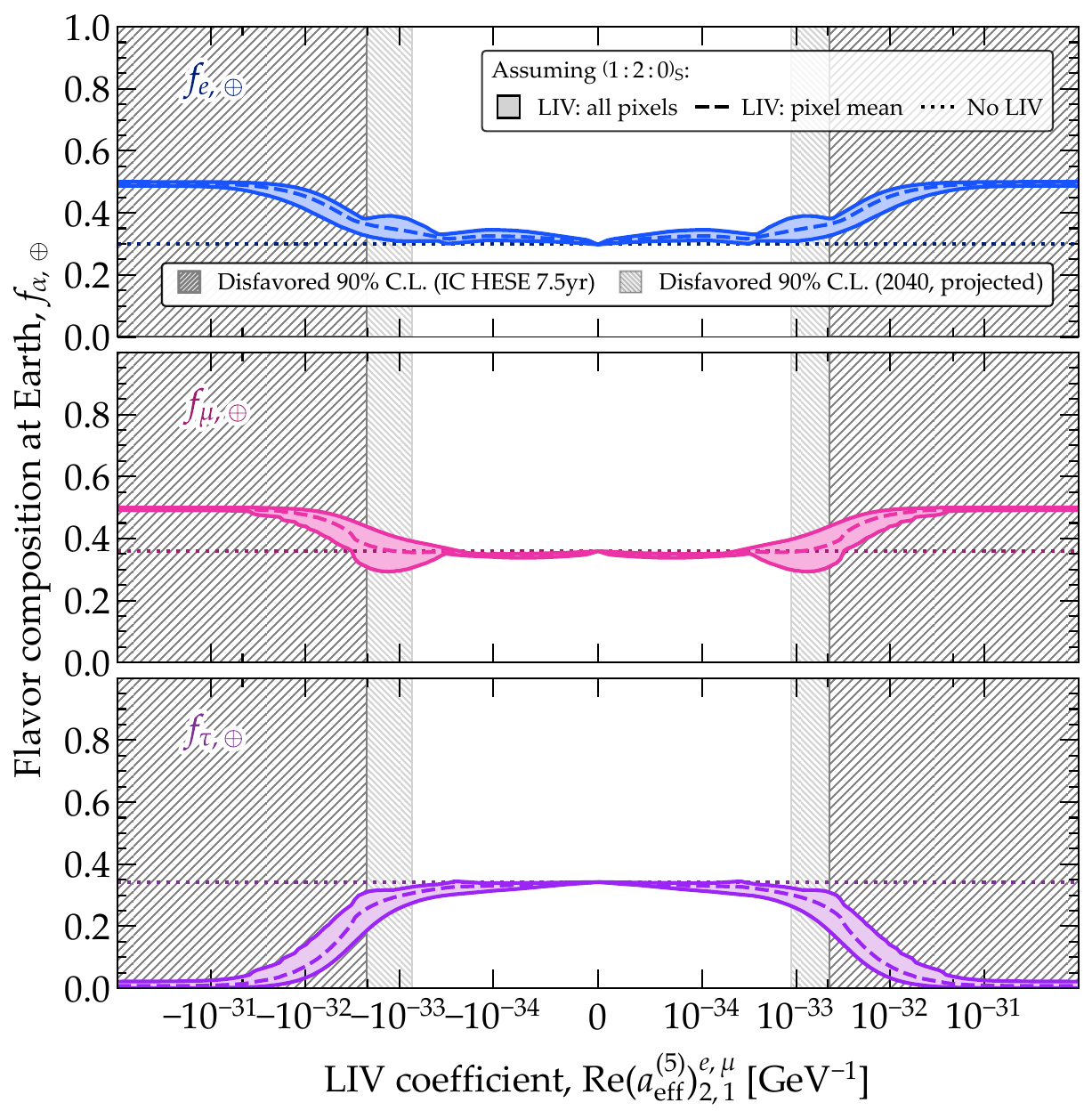}
 \caption{\textbf{\textit{{Limits on the LIV coefficient $\boldsymbol{{\rm Re}(\hat{a}_{\rm eff}^{(5)})^{e,\mu}_{2,1}}$ from the lack of anisotropy in the flavor composition of high-energy astrophysical neutrinos at Earth.}}}  When computing the limits, ${\rm Re}(\hat{a}_{\rm eff}^{(5)})^{e,\mu}_{2,1}$ is the only nonzero LIV coefficient.  The directional variation of the flavor composition introduced by LIV yields the spread in the values of $f_{\alpha, \oplus}$ in the pixels into which the sky is tessellated (``LIV: all pixels''), compared to the standard expectation (``No LIV'').  LIV also shifts the mean all-sky value of the flavor composition (``LIV: pixel mean'') relative to its standard expectation, though our limits are derived from the directional information.  The limits are obtained assuming the canonical expectation of $\left( \frac{1}{3}, \frac{2}{3}, 0 \right)_{\rm S}$, from pion decay, for the flavor composition at the source (see the main text).  Projected limits use the cumulative, combined detected HESE sample from all detectors available by the year 2040.}
 \label{fig:a_emu_21_LIVconstraint}
\end{figure}

Even for the lowest-dimension operator that we consider, $d = 3$, this entails 81 free parameters whose values must be determined.  For the dimension-5 operators that we constrain below, the number is 225, beyond the scope of this first exploration to vary and constrain simultaneously in a full-scale Bayesian approach.  Instead of attempting this, when placing constraints below we turn on only one parameter at a time, the one being constrained, while the rest remain fixed to zero.  This allows us to obtain results within a reasonable time.  Yet, it neglects the possibilities of correlations between parameters and of introducing flavor anisotropy when LIV is dominant (more on this later, in connection to \figu{a_emu_21_LIVconstraint}).

We constrain the value of an LIV parameter, $\lambda$, using the following procedure.  First, we compute the diffuse neutrino flux of each flavor using \equ{propagation}, for many different test values of $\lambda$.  In practice, for each value of $\lambda$, we compute \equ{propagation} at values of $\boldsymbol{\theta}$ given by the center of each of the 12,288 skymap pixels of a HEALPix tessellation with $N_{\rm side}=32$.  Then we downgrade the tessellation to use only 12 pixels, \ie, $N_{\rm side} = 1$, the same resolution that we used earlier to extract the directional flavor composition $f_{e,i}$ and $f_{\mu,i}$ in the main text and in Appendix~\ref{sec:detailed_results}.  (To do this, we use the HEALPix function \texttt{ud\_grade}, which, by coarsening the tessellation, yields an approximate integral over $\boldsymbol{\theta}$ inside each of the 12 pixels.)  Doing so generates, for each value of $\lambda$, skymaps of flavor composition like the one in \figu{LIV_maps_example}.

We place limits on $\lambda$ by comparing these flavor-composition skymaps to our pre-computed posteriors of $f_{e,i}$ and $f_{\mu,i}$.  This allows us to reinterpret our earlier results on directionally dependent flavor composition as constraints on LIV, bypassing the computationally taxing need to generate the event distributions associated to the flux skymaps for each value of $\lambda$ and comparing them to the observed distribution of events.  However, because the posteriors of $f_{e,i}$ and $f_{\mu,i}$ were computed by integrating over all neutrino energies (see the main text), this approach allows us only to extract constraints on $\lambda$ from its effects averaged across all energies.  Accordingly, for each test value of $\lambda$, we average the resulting flavor-composition skymaps in energy, by integrating \equ{propagation} over the approximate IceCube energy range, $10^4$--$10^7$~GeV.  The energy-averaged flavor composition in  pixel $i$ is $f_{\alpha, i}^\prime(\lambda) \equiv \Phi_{\alpha, i}(\lambda) / \sum_\beta \Phi_{\beta, i}(\lambda)$, where $\Phi_{\alpha, i}$ is the flux of $\nu_\alpha$ in that pixel.

In the flavor-composition skymap of $\nu_\alpha$, in each pixel $i = 1, \ldots, N_{\rm pix} = 12$, we compare the composition computed under LIV for a given value of $\lambda$, $f_{\alpha, i}^\prime(\lambda)$, {\it vs.}~the pre-computed value inferred from observation, $f_{\alpha, i}$, via the delta-function likelihood $\mathcal{L}^\alpha_i(\lambda) = \delta(f_{\alpha, i}^\prime(\lambda)-f_{\alpha, i})$.  The posterior distribution of $\lambda$ is the product over all pixels of the $\nu_e$, $\nu_\mu$, and $\nu_\tau$ skymaps, integrated over the pre-computed posteriors of $f_{\alpha, i}$, $\mathcal{P}(f_{\alpha, i})$, \ie,
\begin{eqnarray}
  \mathcal{P}({\lambda}) 
  &=&
  \frac{1}{\mathcal{N}}
  \int df_{\alpha, 1} \ldots df_{\alpha, {N_{\rm pix}}} 
  \prod_{\alpha = e,\mu,\tau}
  \prod_{i=1}^{N_{\rm pix}} 
  \mathcal{L}^\alpha_i(\lambda) 
  \mathcal{P}(f_{\alpha, i}) 
  \nonumber \\
  &=&
  \frac{1}{\mathcal{N}}
  \prod_{\alpha=e,\mu,\tau}
  \prod_{i=1}^{N_{\rm pix}} 
  \mathcal{P}[f^\prime_{\alpha, i}(\lambda)]  \;,
\end{eqnarray}
where $\mathcal{N}$ is a normalization factor such that the posterior integrates to unity.  To speed up the computation, we approximate each posterior $\mathcal{P}(f_{\alpha, i})$ as a gamma distribution fitted to the actual pre-computed posterior.  We report the 90\% credible interval of $\lambda$ by integrating $\mathcal{P}({\lambda})$ around its maximum so that the integral yields 0.9. 

Figure~\ref{fig:a_emu_21_LIVconstraint} shows, as illustration, the resulting limits on the real part of the dimension-5 operator ${\rm Re} (\hat{a}_{\rm eff}^{(5)})_{2,1}^{e,\mu} $, based on present-day IceCube HESE data, $\left\vert {\rm Re} (\hat{a}_{\rm eff}^{(5)})_{2,1}^{e,\mu}  \right\vert < 2.24 \times 10^{-33}$~GeV$^{-1}$, and projected for the year 2040, $\left\vert {\rm Re} (\hat{a}_{\rm eff}^{(5)})_{2,1}^{e,\mu}  \right\vert < 10^{-35}$~GeV$^{-1}$ (90\%~C.L.).  
Broadly stated, using high-energy astrophysical neutrinos improves the reach of limits by $\sim$$10^{15}$ compared to existing limits from GeV accelerator neutrinos~\cite{LSND:2005oop, Kostelecky:2008ts, MiniBooNE:2011pix, Kostelecky:2011gq}.  

Figure~\ref{fig:a_emu_21_LIVconstraint} also illustrates the effect of varying ${\rm Re} (\hat{a}_{\rm eff}^{(5)})_{2,1}^{e,\mu} $ on the directional flavor composition and the three regimes of LIV relative to standard oscillations:
\begin{description}
 \item
  [$H_{\rm LIV} \ll H_{\rm vac}$]
  When $\left\vert {\rm Re} (\hat{a}_{\rm eff}^{(5)})_{2,1}^{e,\mu} \right\vert$ is small, standard oscillations in vacuum dominate, \ie, $H_{\rm LIV} \ll H_{\rm vac}$ in \equ{hamiltonian_tot}, so the flavor composition is isotropic and close to flavor equipartition, $\left( 0.30,0.36,0.34 \right)_\oplus$, as expected from neutrino production by pion decay.
 \item
  [$H_{\rm LIV} \gg H_{\rm vac}$]
  When $\left\vert {\rm Re} (\hat{a}_{\rm eff}^{(5)})_{2,1}^{e,\mu}  \right\vert$ is large, LIV dominates, \ie, $H_{\rm LIV} \gg H_{\rm vac}$.  Again, the flavor composition is isotropic, but different from the expectation of flavor equipartition, since flavor-mixing is now driven solely by LIV.  Isotropy is an artifact of our choice to allow only a single parameter of the $\hat{a}_{\rm eff}^{(5)}$ matrix to be nonzero (see above)---in \figu{a_emu_21_LIVconstraint}, $(\hat{a}_{\rm eff}^{(5)})_{2,1}^{e,\mu}$---which becomes merely a prefactor of $H_{\rm LIV}$.  Because of this choice, the matrix that diagonalizes $H_{\rm tot} \approx H_{\rm LIV}$ is independent of the value of $(\hat{a}_{\rm eff}^{(5)})_{2,1}^{e,\mu}$, and so the resulting flavor composition is isotropic.
 \item
  [$H_{\rm LIV} \approx H_{\rm vac}$]
  In-between the above two regimes, the contributions of LIV and standard oscillations are comparable.  The matrix that diagonalizes $H_{\rm tot}$ is direction-dependent (and energy-dependent), and so is the resulting flavor composition.  In this case, the observation of flavor isotropy places limits on ${\rm Re} (\hat{a}_{\rm eff}^{(5)})_{2,1}^{e,\mu}$ shown in \figu{a_emu_21_LIVconstraint}.
\end{description}



\bibliographystyle{ieeetr}
\bibliography{main.bbl}

\end{document}